**Title:** Microtextures in the Chelyabinsk impact breccia reveal the history of Phosphorus-Olivine-Assemblages in chondrites

**Authors:** Craig R. Walton (1*), Ioannis Baziotis (2), Ana Černok (3), Ludovic Ferrière (4), Paul D. Asimow (5), Oliver Shorttle (1, 6) and Mahesh Anand (3, 7)

1 – Department of Earth Sciences, University of Cambridge, Downing Street, Cambridge CB2 3EQ, UK

2 – Department of Natural Resources Management and Agricultural Engineering, Agricultural Univ. of Athens, Iera Odos 75, 11855 Athens, Greece

3 – Department of Physical Sciences, Open University, Walton Hall, Milton, Keynes, MK7 6AA, United Kingdom

4 – Natural History Museum, Burgring 7, A-1010, Vienna, Austria

5 – Division of Geological and Planetary Sciences, California Institute of Technology, 1200 E California Blvd, Pasadena, CA 91125, USA

6 – Institute of Astronomy, University of Cambridge, Madingley Road, Cambridge, CB3 OHA, UK

7 – Department of Earth Sciences, The Natural History Museum, London, SW7 5BD, UK

*Correspondence to: crw59@cam.ac.uk

**Abstract:**

The geochemistry and textures of phosphate minerals can provide insights into the geological histories of parental asteroids, but the processes governing their formation and deformation remain poorly constrained. We assessed phosphorus-bearing minerals in the three lithologies (light, dark, and melt) of the Chelyabinsk (LL5) ordinary chondrite using scanning electron microscope, electron microprobe, cathodoluminescence, and electron backscatter diffraction techniques. The majority of studied phosphate grains appear intergrown with olivine. However, microtextures of phosphates (apatite [$Ca_5(PO_4)_3(OH,Cl,F)$] and merrillite [$Ca_9NaMg(PO_4)_7$]) are extremely variable within and between the differently-shocked lithologies investigated. We observe continuously strained as well as recrystallized strain-free merrillite populations. Grains with strain-free subdomains are present only in the more intensely shocked dark lithology, indicating that phosphate growth predates the development of primary shock-metamorphic features. Complete melting of portions of the meteorite is recorded by the shock-melt lithology, which contains a population of phosphorus-rich olivine grains. The response of phosphorus-bearing minerals to shock is therefore hugely variable throughout this monomict impact breccia. We propose a paragenetic history for P-bearing phases in Chelyabinsk involving initial phosphate growth via P-rich olivine replacement, followed by phosphate deformation during an early impact event. This event was also responsible for the local development of shock melt that lacks phosphate grains and instead contains P-enriched olivine. We generalise our findings to propose a new classification scheme for Phosphorus-Olivine-Assemblages (Type I-III POAs). We highlight how POAs can be used to trace radiogenic metamorphism and shock metamorphic events that together span the entire geological history of chondritic asteroids.

# INTRODUCTION

## Phosphorus in meteorites: knowns and unknowns

Meteorites provide direct samples of some of the most primitive solid materials found in the Solar System, yielding insights into early disk processes including chemical partitioning in the protoplanetary disk, the assembly of dust into planets, and subsequent dynamical evolution (e.g., Scott, 2007). We now have a broad understanding that the parent asteroids of chondritic meteorites experienced heating (recorded by thermal metamorphism) predominantly by the decay of short-lived radionuclides for the first 60 million years (Myr) of Solar System history (e.g., Bouvier et al., 2007). A large body of evidence shows that impact events continued to disturb these objects beyond the end of thermal metamorphism, resetting a number of mineral geochronometers in the process (e.g., Wittmann et al., 2010). Both radiogenic and impact-induced heating are thought to mobilize fluids within chondrites (Zhang et al., 2016; Lewis and Jones, 2016), resulting in redistribution of volatile components and growth of secondary phases. Thus, the elemental and isotopic compositions of volatile-bearing phases in meteorites and other planetary samples have the potential to provide important insights into both planetary volatile reservoirs and processes that operated on their parent bodies (e.g., Jones et al., 2014; Černok et al., 2020; Stephant et al., 2019). Volatile-bearing phases that retain such information include halides (Jones et al., 2016), sulfides (Visser et al., 2019), and phosphates (Jones et al., 2016).

In this context, studies of phosphorus (P)-bearing minerals have been especially informative. In the most primitive unequilibrated chondrites (e.g., carbonaceous chondrites), P is found dissolved in Fe-Ni metal as a neutral minor element (Zanda et al., 1994), indicating siderophile behavior during early condensation. Indeed, thermodynamic models predict that P initially condenses from the solar nebula as schreibersite [$(FeNi)_3P$]. Schreibersite is often observed to be non-stoichiometric, with P < 1 atom per formula unit (Zanda et al., 1994; Pasek, 2019). Conversely, in thermally metamorphosed and aqueously altered chondrites, P is found in its oxidized form ($P^{5+}$) within the phosphate minerals merrillite [$Ca_9NaMg(PO_4)_7$] and apatite [$Ca_5(PO_4)_3(OH,Cl,F)$], implying that P is scavenged from metal and phosphide grains by fluids and redistributed within the parent body (Jones et al., 2014).

The direct link to volatile components arises because apatite requires structural volatile components (OH, Cl, F) in order to crystallize. Apatite and merrillite also incorporate U into their crystal lattices, making both phases viable for U–Pb dating (e.g., McGregor et al., 2019; Merle et al., 2014; Yin et al., 2014) and hence for time-resolved studies of volatile component evolution. Phosphate U–Pb dating in chondrites has helped to map out radiogenic heating on million-year timescales, as well as near instantaneous resetting during shock-metamorphic events (e.g., Yin et al., 2014). It is therefore possible to tie the history of asteroidal phosphates to the compositional evolution of metasomatic fluids during both early and late-stage heating events on asteroidal parent bodies. Additionally, a third mineral reservoir of P has recently been identified in chondrites: P-enriched olivine grains of enigmatic origin (e.g., Li et al., 2017; McCanta et al., 2016).

It is clear that P-bearing minerals have a role to play in elucidating the complex histories of asteroidal parent bodies. However, there are numerous unresolved issues both in the study of P-bearing phases and asteroidal evolution in general – many of which would appear to be linked. In particular, it remains unclear (1) how oxidized P-bearing minerals form in chondrites from initially reduced Fe-bound P (Jones et al., 2014, 2016), and (2) how these P-bearing phases respond to impact-processing (Cox et al., 2020; McGregor et al., 2018; Krzesińska et al., 2017). Given widespread evidence for shock metamorphism in meteorites, unravelling which aspects of the phosphate record pertain to radiogenic versus impact-induced metamorphism is critical for a full understanding of the physicochemical evolution of asteroids.

Here, we assess both, the phase associations and microtextures of phosphate minerals and P-bearing olivine in a highly shocked chondritic meteorite, i.e., Chelyabinsk. We use a combined imaging approach in order to access records of heterogeneous shock processing throughout the meteorite. By comparing our results with published literature, we are able to establish new paragenetic constraints on P-hosting phases – from their initial formation, through subsequent impact-induced metamorphism and melting.

**Phosphorus in the Chelyabinsk impact-breccia: an ideal case-study?**

Chelyabinsk is an LL5 (low-metal, low iron, petrologic type 5 — i.e., thermally equilibrated) S4-6 (moderately shocked to shock-melted; Morlok et al., 2017; Fritz et al., 2017; Stöffler et al., 1991) impact-brecciated meteorite. The Chelyabinsk impact breccia contains three distinct shock-related lithologies, which are distinguished at a first-order level by their optical color: (1) a light-colored, chondritic-textured lithology with shock melt veins (SMVs) and high-pressure phases; (2) a dark-colored lithology with relict shock-darkened chondritic textures, abundant shock melt pools, and SMVs with high-pressure phases; and (3) a quenched shock-melt lithology with some entrained chondritic fragments and an absence of high-pressure phases (Morlok et al., 2017; Righter et al., 2015; Koroteev et al., 2013; Popova et al., 2013; Bischoff et al., 2013).

Despite the extensive work already performed on Chelyabinsk, the textural setting of P-bearing minerals with surrounding phases, their origin and internal microtextures have not yet received specific attention. Both the light and dark lithologies of Chelyabinsk contain apatite and merrillite (e.g., Popova et al., 2013) yet these phases are conspicuous for their absence in the shock-melt lithology. The range of lithologies preserved in Chelyabinsk may contain assemblage-scale textures and mineral-scale microtextures that pertain both to the initial growth and to the subsequent impact metamorphism of chondritic phosphate minerals. The great diversity of petrological, geochemical, and geochronologic analyses that have been reported for each lithology provide a broader context within which to construct self-consistent models of phosphate paragenesis.

Previous studies have gained insight into phosphate genesis and response to impact by evaluating associations and microtextures of mineral phases that tend to co-exist together or in close proximity to phosphates (e.g., Jones et al., 2014). Here, we present a thorough analysis of P distribution in phosphate and olivine grains within all three of the variously shocked lithologies found in the Chelyabinsk meteorite. We consider the implications of observed mineral microtextures, phase associations, and chemical compositions for the relative timing of thermal and shock metamorphic events affecting phosphorus reservoirs in Chelyabinsk, concluding with a discussion of newly observed microtextures that demand attention during future studies of shocked meteorites.

## MATERIALS AND METHODS

### Samples

Several polished sections of Chelyabinsk from the collections of The Open University (OU) (sections 'A' and 'B') and the Natural History Museum (NHM), Vienna (NHMV; N9834) were investigated during the course of this study. Section 'A' contains the light lithology of Chelyabinsk (Pillinger et al., 2013), a moderately shocked (dominantly S4) chondrite with well-preserved chondrules and matrix, crosscut by SMVs. Section 'B' contains the dark lithology (Pillinger et al., 2013), more severely shocked (blackened) material and numerous SMVs (dominantly S5). Finally, section N9834 contains regions of shock-blackened chondrite (S5) as well as quenched shock-melt (S6).

**Light lithology**

The light lithology of Chelyabinsk is dominated by thermally metamorphosed and chemically equilibrated matrix and chondrule material, in which major and minor mineral phases include silicates (olivine, pyroxene, feldspar), troilite (FeS), iron-nickel (P-free Fe-Ni) metal, phosphates (apatite and merrillite), and chromite (Righter et al., 2015). SMVs provide clear evidence for the effect of shock metamorphism in the light lithology, but typically comprise only several volume percent of a given sample. SMVs have bulk compositions broadly representative of the whole-rock composition and preserve evidence of rapid cooling (e.g., micro-to-cryptocrystalline silicates, metal-troilite globules, and glasses; Kaeter et al., 2018; Righter et al., 2015).

Shock features are also pervasive in the chondrite-textured portion of the light lithology. For example, phases with high melting temperatures, such as olivine and pyroxene, are generally fully crystalline yet exhibit planar fractures, shock-induced mosaicism, as well as evidence of shearing (e.g., displacement across micro-faults; Righter et al., 2015; Kaeter et al., 2018). Phases with lower melting temperatures — such as plagioclase, metal, and troilite — have widely formed veins through un-melted silicates (Kaeter et al., 2018).

Phase-specific responses are more complicated on smaller scales, where local density contrasts between phases induce locally heterogeneous melting behavior (Moreau et al., 2018). There are examples of metal and troilite grains that retain their pre-impact textures, as well as examples that have clearly experienced mobilization during the impact (e.g., injection along grain-boundaries and into fractures, forming network and vein textures; Korotev et al., 2013). Meanwhile, all plagioclase feldspar would appear to be secondary, having experienced melting followed by quenching or recrystallization (Kaeter et al., 2018) and for which a number of different occurrences have been identified (e.g., plagioclase-normative glass, maskelynite, plagioclase-chromite symplectites, crystalline plagioclase feldspar; Kaeter et al., 2018). Where possible, we use the same terminology as Kaeter et al (2018) to describe feldspathic phases (i.e., depending on available observations). Otherwise, the generic term plagioclase is used.

**Dark lithology**

The dark lithology is mineralogically similar to the light lithology. However, the chondrite-textured portion is more heavily brecciated. Sulfide, metal, silicate-rich SMVs, and shock-melt pools constitute a greater volume percentage of dark samples. Planar deformation features in olivine are much more extensive. These features contribute to a pervasive and diagnostic shock darkening (Kaeter et al., 2018; Righter et al., 2015; Krzesińska et al., 2015; Rubin, 2003a,b). Plagioclase is more widely associated with chromite in symplectite structures (interpreted as former melt pools).

**Shock-melt lithology**

The shock-melt lithology is a micro-to-cryptocrystalline breccia composed mostly of olivine-pyroxene-glass mesostasis along with fragments of chondrite-textured material (both light and dark lithologies) and metal-troilite droplets (Righter et al., 2015).

**Microtextural characterization of phosphates and olivine**

Both the light and the dark lithology samples of Chelyabinsk from the OU collection (Pillinger et al., 2013) were characterized using Scanning Electron Microscopy (SEM), Zeiss Supra 55VP at the OU and Quanta 650 at the University of Cambridge. Whole-sample Backscattered Electron (BSE) images and elemental maps using Energy Dispersive X-ray Spectrometry (EDS) were collected for each sample in order to identify the distribution of key phases and provide context for higher-magnification analyses. Merrillite and apatite were distinguished on the basis of EDS measurements in point mode, with 10 second spectral acquisitions, based on the abundance of Na, Mg and Cl. Additional SEM imaging of the shock-melt lithology was performed at the California Institute of Technology (Caltech) Geological and Planetary Sciences Division Analytical Facility using a Zeiss 1550VP field-emission SEM equipped with an angle-sensitive backscattered electron detector, 80 $mm^2$ active area Oxford X-Max Si-drift-detector EDS, and an HKL EBSD system. The SEM imaging and EDS analyses used a 15 kV accelerating potential and a 120 µm field aperture in high-current mode (~4 nA probe current), yielding imaging resolution better than 2 nm and an activation volume for EDS analysis ~1–2 $\mu m^3$ on silicates.

SEM-Cathodoluminescence (CL) images were collected on selected phosphate and plagioclase grains previously imaged in BSE, in order to obtain an initial view of crystal structure integrity across the sample. The CL images were acquired on the OU SEM with a Deben Centaurus CL panchromatic detector with Hamamatsu Photo Multiplier Tube (model R316), with spectral response to wavelengths from 400–1200 nm. Imaging conditions were 7–10 kV accelerating voltage, 12–13 mm working distance, and high vacuum mode.

We quantify phosphate phase associations in Chelyabinsk by counting the phases that share a grain boundary with a phosphate mineral. Phases associated with phosphate in this study include olivine, pyroxene, plagioclase, metal, and sulfides (Fig. 3).

Lattice orientation, internal microtexture, and structural disorder of selected phosphate minerals were studied by EBSD. For the light and dark lithologies, after SEM analyses, isopropanol and, if needed, 0.25 μm diamond paste were used for removing the carbon coat from the samples. Subsequently, the thin sections were lightly polished for approximately 10 minutes using either 50 nm alumina or colloidal silica in water suspension with a LabPol-5 system with a LabForce-1 head and an automated doser at OU, to ensure removal of surface defects. This process minimizes scatter of the electron beam due to surface interaction and is a critical step for accurate analysis by EBSD.

We constructed images from rasters of EBSD solutions using an Oxford Instruments Nordlys EBSD detector mounted on the OU SEM. Diffracted electrons were collected at a tilt angle of 70º. Raster images were constructed with 400 to 550 nm step sizes and collection times < 120 ms per step. Binning on the EBSD area detector was set to 4 x 4 μm (e.g., White et al., 2017; Černok et al., 2019). Voltage was set to 20 kV with a largest aperture of 120 microns and high current mode. Such conditions typically generate a beam current of 9.1 nA, measured in a Faraday cup at the same conditions. The interaction volume for diffracted electrons using these parameters is estimated to be a few tens of nm in width and depth (Darling et al., 2016). Raw patterns are automatically background corrected during acquisition, removing an averaged 64 frames of 'noise' from each pattern. Wild spike reduction and minimal zero solution correction were the only raw data corrections applied. Collected diffraction patterns were indexed either as apatite, with the hexagonal unit cell parameters $a$=9.4555, $b$= 9.4555, $c$= 6.8836, $\alpha$=90º, $\beta$=90º, and $\gamma$=120º (Wilson et al., 1999) or as merrillite with the trigonal unit cell parameters $a$=10.3444, $b$=10.3444, $c$= 37.0182, $\alpha$=90º, $\beta$=90º, and $\gamma$=120º (Xie et al., 2015).

For the shock-melt lithology, single crystal EBSD analyses of olivine at a sub-micrometer scale were performed at 20 kV and 6 nA in focused beam mode with a 70° tilted stage on uncoated specimens in "variable pressure" mode (25 Pa of $N_2$ gas in the chamber to reduce specimen charging). Imaging, mapping, semi-quantitative EDS analysis, and EBSD were conducted using the SmartSEM, AZtec, and Channel 5 software packages, respectively.

Crystallographic orientation of individual grains and their internal microtexture obtained by electron diffraction can be visualized in different ways. Band Contrast (BC) is a measure of the pattern quality of a material, based on the contrast between observed Kikuchi bands and the background. This measure is affected by phase crystallinity, lattice orientation, defect density, surface polishing conditions, and SEM/EBSD set-up parameters (Kang et al., 2013). Texture component (TC) maps reveal orientation of individual grains and subdomains within grains, and reveal distortion of the crystal lattice from a reference point. Inverse Pole Figure (IPF) maps show absolute orientation of the normal to the polished surface in the crystal lattice coordinate system and help visualize the orientation of the lattice of the entire grain. In particular, TC and IPF maps help distinguishing strained from unstrained grains and subdomains.

**Phase association quantification**

We quantified phosphate phase associations by manually counting the grain boundary contacts shared by each phosphate grain studied with neighboring phases. This form of quantification results in an overall percentage of grains that share a contact with e.g., olivine.

**Electron microprobe**

We studied the composition and zoning of phases in point and mapping modes using a JEOL JXA-8530F Field Emission electron microprobe (EPMA) at the NHMV and a Cameca SX100 EPMA at the OU. All analyses were performed with an accelerating voltage of 15 kV. For minerals, a 20 nA focused beam current, 20 s counting time on peak position and 10 s for each background were used. Selected sites were re-analyzed for P at 50 nA beam current with 20 s peak and 10 s background counting time. Detection limit (1 sigma) for P was ~ 76 ppm. For glass analyses, a slightly defocused (5 µm diameter) beam and 10 s counting time were used. Natural mineral standards used were albite (Na, Si, Al), wollastonite (Ca), olivine (Mg), almandine (Fe), spessartine (Mn), orthoclase (K), rutile (Ti), chromite (Cr), Ni-oxide (Ni), and apatite (P).

## RESULTS

Whole-section BSE image mosaics are shown in Figure 1. Areas studied at higher resolution in each section (and which are discussed in the main text) are called out with subframe rectangles or grain numbers. Individual phosphates are named according to their image number, phase identity, and section number e.g., 001-AP-A would be an apatite from image 001 of section A (our sample of the light lithology), whereas 001-MERR-B corresponds to a merrillite from image 001 of section B (our sample of the dark lithology). Silicate grains from the shock-melt lithology are named according to the sequence and analytical session in which they were studied.

### Phosphorus phase associations

*Light and dark lithologies*

Phosphates in both the light and dark lithologies of Chelyabinsk occur mostly as individual grains, as either apatite or merrillite. Only a few examples were found of close association between the two phases (occurrences as replacement or intergrowth). Phosphate grain boundaries truncate against other phases sharply except in the case of olivine, where contacts are often irregular and complex/embayed (Fig. 2A). Both apatite and merrillite, across all textural settings and associations, were observed to contain small (<10 μm) rounded inclusions of olivine (Fig. 2A, label 2), as well as, more rarely, plagioclase, chromite, sulfide, and pyroxene. Phosphate inclusions are often observed to occur in olivine (Fig. 2A). Some small (<10 μm) phosphate and phosphate-olivine inclusions also occur in pyroxene (Fig. S3).

SMVs are occasionally observed to truncate or intrude a number of features, including phosphate grains (Fig. 2B). Phosphates and olivine display evidence of having interacted with mobile plagioclase, including thin rims around and linear fills of apparent planar deformation features in olivine and pyroxene, as well as zones marked by finely interspersed olivine and chromite in plagioclase (Fig. 2C). Some phosphate grains were encountered in close association with Fe-Ni metal (Fig. 2D) and chromite grains, but the majority are in contact with silicate phases.

Our quantitative assessment of phosphate phase associations reveals that in the light lithology both apatite and merrillite are preferentially associated with olivine and plagioclase, with significantly fewer crystals sharing boundaries with pyroxene, sulfide, or metal (Fig. 3 – raw data available in Table S2). On the other hand, associations are much less organized and more complex in the dark lithology, with the majority of phosphates associated with all five other phases considered (Fig. 3).

*Melt lithology*

We do not observe phosphates in the shock-melt lithology. The budget of P in the shock-melt lithology is instead hosted as a minor element in silicate phases. Our sample of the shock-melt lithology contains around 50 vol.% quenched impact-melt, composed of small (< 5 μm) zoned olivine and orthopyroxene crystals, silica-rich glass, and finely-disseminated sulfide grains. Relict chondrules and individual grains of olivine, pyroxene, and plagioclase entrained in the melt are highly resorbed and surrounded by melt. Quench textures include a high abundance of melt inclusions in silicate phases, along with concentric-to-interstitial skeletal growths of olivine. Compositional gradients are apparent in the melt in proximity to resorbed mineral phases.

Orthopyroxene occurs as large porphyroclasts (>150 μm in length), as small subhedral to anhedral crystals (15–20 μm in length), and as microlites (~10 μm in length) entrained in the melt. The large porphyroclasts have a composition $En_{75-78}Fs_{22-24}$; the small crystals have a broader range of composition $Wo_{0-5}En_{68-86}Fs_{14-27}$; the microlite crystals show a homogeneous Mg-rich composition $En_{87}Fs_{13}$ (Table S2B). The $P_2O_5$ abundances are low (≤ 0.06 wt%) in all orthopyroxene grains, with the exception of the microlite crystals, which exhibit a range from 0.13–0.19 wt% (Table S2B). Note that the P enrichment in microlite orthopyroxene crystals is not an artifact of contamination from neighboring phases, which have lower P content than the orthopyroxene itself.

EPMA defocused spot analyses on pools of glass larger than 10 μm in diameter reveal a $SiO_2$-rich composition (63.3–64.9 wt%) with very low $P_2O_5$ content ≤ 0.05 wt% and Mg# in the range of 24.4–40.8 (Table S2C). Unfortunately, the small glass areas between quenched crystals and the zoned rims around resorbed phases are too small for clean defocused beam EPMA analyses, so the compositional range given here may not be fully representative.

Our sample of the impact melt contains olivine grains ($Fo_{64.5-85.8}$) with 0.02–0.52 wt% $P_2O_5$ (Table S2C). Two populations of olivine in this lithology are distinguished on the basis of textures, major element chemistry, and P abundance: partially resorbed relics and quench olivine, found as individual crystallites and rim-overgrowths around partially resorbed relics (Fig. 4). Quench olivine is clearly zoned as visible on BSE-SEM images (Fig. 4A). Quench olivine is further subdivided into individual crystals and rim-overgrowths that formed around pre-existing olivine crystals. Figure 4B illustrates a schematic view of these various olivine populations.

Individual quench and rim-overgrowth quench olivine share geochemical and textural similarities. Both olivine types are concentrically zoned, with the inner Mg-rich portion containing a high proportion of silicate glass inclusions. The inner zones of both individual and rim-overgrowth crystals are also somewhat P-enriched compared to any preserved core olivines (Fig. 5). The outer portion is generally inclusion-free, Fe-rich, and very P-rich (Fig. 5). Phosphoran olivine is defined by having > 1 wt% $P_2O_5$ (Boesenberg and Hewins, 2010). Our samples display a maximum of ~0.5 wt% $P_2O_5$ and are hence most accurately referred to simply as variously P-rich olivine.

Figure 6 presents a scatter plot of FeO vs. $P_2O_5$ contents for the various olivine populations identified in the different lithologies of Chelyabinsk. Quench olivine data are divided by backscatter contrast into dark (typically the inner regions) and light (typically near the rims) groupings. High backscatter-contrast outer parts of quench olivine in the melt lithology have P concentrations substantially higher than in resorbed cores, the inner parts of quench olivine, or olivine in the light lithology. Phosphorus concentration in high backscatter-contrast outer parts of olivine rims is also higher than that in any other phase in the quenched shock-melt lithology, including glass and pyroxene (Table S2).

We rule out that the P-in-olivine signature merely represents contamination by silicate mesostases on the basis of both visual (Fig. 5) and geochemical (Fig. 6) evidence. All of the high P concentrations (≥0.25 wt% $P_2O_5$) measured in olivine are in outermost light-colored quench rims (Fig. 5), which are generally inclusion-free (see spot overlay; Fig. S6; full silicate chemistry data are provided in Table S2). We further rule out that the P-in-olivine signature merely represents contamination by silicate mesostases on the basis of a positive correlation between FeO and $P_2O_5$ in the quench olivine population (Fig. 6). These data support a model in which P is directly incorporated into the olivine structure, as contamination by silicate mesostases would yield lower FeO.

The rims of quench olivine in Chelyabinsk exhibit moderate positive correlations between $P^{5+}$ and trivalent cations ($Al^{3+}$ + $Cr^{3+}$; Fig. 7A) and between $P^{5+}$ and a deficit relative to 3 cations per 4-oxygen formula unit (R=0.64; Fig. 7B). In contrast, P is poorly correlated with $Si^{4+}$ (R=0.28; Fig. 7C). Based on these correlations, we infer that P substitution was predominantly in exchange with divalent cations, alongside trivalent Cr and Al, and balanced by M site vacancies, i.e., $4^{VI}M^{2+}={}^{VI}(Cr,Al)^{3+}+{}^{VI}P^{5+}+{}^{VI}[]$. Core olivine compositions are identical to the typical equilibrated olivine found in both the light and dark lithologies (Fig. 6).

## Mineral microtextures

Microtextural EBSD and CL data provide information with which we test predictions made by different models of P-mineral paragenesis. CL imaging is sensitive to surface and internal mineral textures, structural defects, trace element distribution, and elemental zoning in plagioclase and phosphates (Götze and Kempe, 2008). Changes in phase crystallinity and element mobility during impact metamorphism may each create visible signatures in CL images, but which are more reliably interpreted when provided with context from other techniques, such as EBSD. We studied the microtextures of P-bearing minerals, as well as other closely associated phases, in order to determine a self-consistent paragenetic history.

An important novel finding arising from our observations is that phosphate microtextures differ systematically both among lithologies with different shock stages, and within each lithology owing to the local phase assemblage. We identify plastically-strained apatite and merrillite grains in the light lithology, whereas in the dark lithology we observe strained apatite grains and recrystallized merrillites. We also find a novel patchy phosphate population in both the light and dark lithologies. Figures 8-10 present analyses performed on phosphates. Figure 11 presents EBSD data for olivine in the light, dark, and melt lithologies. Figure 12 summarizes in schematic form the most important phosphate textures observed via EBSD, CL, and EDS imaging in the course of this work.

*Plastically-strained versus recrystallized phosphates*

We identify a dichotomy in phosphate textures between the differently shocked lithologies of Chelyabinsk: plastic deformation of phosphates in the light lithology versus recrystallized phosphates (exclusively merrillites, given available data) with strain-free subdomains in the dark lithology. Strained grains show plastic deformation across the grain that results in total cumulative misorientations of up to 16°, as visible in TC and IPF maps and respective pole figures (Fig. 9A-B, and related pole figures). In contrast, the strain-free subdomains we observe in recrystallized merrillite show significant misorientations from one another as demonstrated by TC and IPF maps (Fig. 9C), but retaining obvious memory of the original parent crystal orientation, as visible by the dominant orientation pattern in pole figures (Fig. 9C). Meanwhile, individual subdomains demonstrate no internal strain.

Recrystallization textures are visible in both CL images (Figs. 8-9) and EBSD data (Fig. 9), but not in EDS X-ray maps (Fig. 10). The merrillite subdomain orientations shown in Figure 9C appears dominantly orthogonal to the orientation of the neighboring SMV. The EBSD maps reveal that individual subdomains are highly crystalline (returning patterns of sufficient quality to index at every pixel), whilst CL images demonstrate complex patchy areas within each crystal subdomain. In another recrystallized merrillite, outer subdomain and grain boundaries are marked by notably bright CL responses (Fig. 8C). Bright spots also occur in grain interiors (Figures 8C and 9C). Recrystallization textures are universally well-developed in dark lithology merrillites but are apparently absent in dark lithology apatites (Fig. 8B).

In contrast, all phosphates in the light lithology lack subdomain formation and instead display evidence of internal plastic deformation (Figs. 8A, 9A-B). TC images of these phosphates reveal moderate crystal-plastic deformation of the grains with up to ~16° of internal misorientation from the reference point, with no evidence of recrystallization (as supported by IPF maps and relevant pole figures; Fig. 9A).

*Patchy phosphate cathodoluminescence signature*

In both the light and the dark lithologies, some apatite grains show complex CL response, displaying generally thin (< 10 μm) patches of lower CL intensity that may either follow fractures and grain boundaries or appear to be independent of them (Figs. 8A-B and 9A). This patchy CL response is visible also in some recrystallized dark lithology merrillite grains (Fig. 8C) but is apparently absent in light lithology merrillites (Figs. 8A and 9B). There is no obvious overlap between the patchy textural features we observe in CL and the microtextures revealed by EBSD (Fig. 9). The EDS X-ray maps also do not reveal any visible features in elemental chemistry that would correspond to the patchy CL textures that we observe.

*Variable olivine microtexture*

The microtexture of olivine in the shock-melt lithology (Fig. 11) contrasts markedly with that found in either the light or dark lithology (Fig. 11A-D). Individual olivine grains in the light lithology show minimal internal strain (Fig. 11A). Olivine in the dark lithology displays local misorientations of a few degrees, which is indicative of weak plastic deformation (Fig. 11B). Olivine in the melt lithology displays a wider diversity of textures. We observe ~100 µm relict olivine cores that are divided into mosaics of randomly oriented 5-10 µm subdomains (i.e., recrystallization; Fig. 11C). Also observed are quench growths of olivine which have large strain-free subdomains, which yield tightly clustered pole figures (Fig. 11D). Both the relict recrystallized and quench olivine populations display good crystallinity and no evidence of internal misorientation (Fig. 11C-D).

## DISCUSSION

### Genesis of P-bearing minerals

*Phosphate formation: metasomatism or shock melting?*

The complex microtextures and textural associations of phosphates in the Chelyabinsk meteorite record key events in the history of the meteorite and its parent body. A key observation is that Chelyabinsk phosphates are predominantly in contact with olivine and plagioclase (Figs. 2 and 3). Our observations contrast with Lewis and Jones (2016) and Jones et al. (2014), who reported that phosphates in unshocked L and LL chondrites showed no preferred association with any particular silicate phase. Rather, phosphates in those lower shock stage samples were nearly equally associated pyroxene, plagioclase, and olivine, being most often found in phase assemblages that contain all of these minerals (Lewis and Jones, 2016; Jones et al., 2014).

Moreover, Jones et al. (2014) and Lewis and Jones (2016) identified (but did not quantify) common association of phosphates with Fe-Ni metal and sulfide grains, as well as voids resulting from plucking of Fe-Ni metal and sulfide grains during sample preparation. However, light lithology phosphates in Chelyabinsk almost never occur in such an association (Fig. 3; ~5 % association with voids, see Table S1). Therefore, the distribution of phosphates in Chelyabinsk is anomalous among ordinary chondrites. Noting that Chelyabinsk is unremarkable in terms of its bulk chemical composition and modal mineralogy among the LL chondrites (Sharygin et al., 2013), we consider several hypotheses to explain our observations.

One obvious difference is that Chelyabinsk is heavily shocked, whereas Lewis and Jones (2016) and Jones et al. (2014) focused on relatively unshocked ordinary chondrites. However, there remain a number of aspects of the phosphate distribution in Chelyabinsk that are similar to textures identified by Jones et al. (2014). Previous studies have identified phosphate-olivine reaction rims suggesting the growth of phosphate by replacement of olivine during aqueous alteration (Lewis and Jones, 2016; McCubbin and Jones, 2015; Jones et al., 2014). The intergrowth textures present in Chelyabinsk (e.g., Fig. 2A) could therefore be interpreted similarly to those in relatively unshocked ordinary chondrites i.e., as partial replacement of olivine by phosphate minerals.

Alternatively, the phosphate-olivine intergrowths found in Chelyabinsk may form upon crystallization from shock-melts (Jones et al., 2014; Krzesińska, 2017). Phosphate is readily consumed during impact melting of chondrites (Donaldson, 1985; Brunet and Laporte, 1998; Chen and Zhang, 2008). In a shock-melt model for silicate-phosphate intergrowths, pre-existing olivine-phosphate intergrowths that developed during parent body thermal metamorphism melt and then regrow as new olivine-phosphate intergrowths. Supporting this idea requires detailed examination of the textures of the olivine-phosphate intergrowths, for features that uniquely develop upon growth from melts at high cooling rates or degrees of undercooling. Distinctive distributions of P in olivine, for example, can be found in unaltered igneous rocks and rapidly cooled crystallization experiments (e.g., Milman-Barris et al., 2008). In the case of Chelyabinsk, this would manifest as anomalously P-rich compositions for olivine found in symplectite-like phosphate-olivine assemblages.

The mineral associations and microtextures that we describe here help to distinguish between these different possibilities. In summary, we evaluate two possibilities for the origin of phosphates in Chelyabinsk: (1) growth via metasomatism during thermal metamorphism on its parent body (an early Solar System process), and/or (2) growth as the crystallization products of impact-induced shock-melts.

*Evidence from the light and dark lithologies*

The two formation mechanisms we have outlined are not necessarily mutually exclusive. However, our results allow us to rule out phosphate formation during the impact event that formed the Chelyabinsk impact breccia.

First, the finely crystalline quench textures of melt veins appear to post-date (i.e., crosscut) the relatively coarse intergrowth textures described for most phosphate-olivine textures in the light lithology of Chelyabinsk (e.g., Fig. 2B). Frictional melting along planes of displacement (now preserved as SMVs) has therefore affected some of the phosphates in Chelyabinsk. At the same time, this observation requires that the phosphates existed prior to the impact event – rather than having formed during it.

During this event, light lithology phosphates became deformed and dark lithology merrillites were recrystallized (e.g., Figs. 7, 8, and 10). The recrystallization textures of dark lithology merrillites are pristine (i.e., unstrained subdomains), showing that there has been no significant metamorphic overprint or alteration that postdates the shock melting event that produced the Chelyabinsk breccia. Undeformed igneous grains show tightly clustered pole figures in EBSD data. Pole figures of recrystallized merrillites in Chelyabinsk instead show evidence for memory of parent crystal orientation in the form of extended arcs: this is unambiguous evidence that these grains did not grow from an impact-induced melt (Fig. 9C). Therefore, the phosphate minerals found in those lithologies must also predate the shock-melting impact during which they recrystallized.

Phosphates in the light lithology display significant crystal-plastic deformation (up to 16° internal misorientation) yet preserve minimally deformed olivine inclusions with similar crystallographic orientation (Fig. 8, Panel C2; Fig. S6). Both of these observations clearly imply that these phosphate-olivine intergrowths must predate the impact that deformed them. Our observations suggest the growth of coherent phosphate grains, potentially via olivine replacement, to leave islands of disconnected but similarly oriented primary olivine.

Dark lithology phosphates also contain large rounded olivine inclusions. However, in this instance SMV-proximal olivine inclusions display complex and strained microtexture (Fig. 9, Panel C3, and Fig. S5). We also observed cases where recrystallized phosphates and their olivine inclusions have been cut by and displaced by sliding across SMVs (Fig. 8). Finally, we find that P concentrations are uniform in olivine throughout the light lithology (Fig. 6). This finding suggests that olivine intergrown with phosphate in the light and dark lithologies did not grow from a P-rich melt. All of these observations suggest that metasomatic replacement of olivine by phosphate occurred before impact metamorphism.

*Evidence from the melt lithology*

The shock-melt lithology lacks any discrete phosphate phases. This observation suggests the complete melting of phosphates during impact, followed by formation of P-rich olivine during quench-crystallization of the resulting melt. A quench origin is obvious based on the petrographic context, quench textures (e.g., high abundance of melt inclusions, concentric-to-interstitial skeletal growths; Fig. 4) and the anomalous chemistry of these olivine grains (Fig. 5), including boundary enrichment of phosphorus.

The EBSD maps of light, dark, and melt lithology olivine contrast significantly. In the light lithology, olivine is only very mildly deformed (Fig. 11A). In the dark lithology, olivine shows weak-moderate level of crystal-plastic deformation when in direct contact with SMVs (Fig. 11B). In the shock-melt, relict cores are recrystallized, and both these and quench olivines are apparently strain-free (Fig. 11C-D). Evidently, both the quench-crystallized P-rich and relict P-poor recrystallized olivine, like the recrystallized phosphates, have not experienced additional deformation since impact-related (de)formation.

Recrystallized relict olivine cores in the shock-melt served as nuclei for growth of the subhedral to euhedral branches (Fig. 4; see Erdmann et al., 2014 for similar textures). Indeed, the texturally distinct core and quench olivine in the shock-melt lithology also display contrasting chemical compositions (Fig. 5). Combined with increasing backscatter brightness towards the rim, later crystallizing quench olivine contains higher concentrations of elements that are generally incompatible in the crystal structures of olivine and pyroxene (e.g., P), i.e., those silicate phases observed to have freshly crystallized in the shock melt lithology. This relationship could be simply interpreted as reflecting growth of later crystallizing silicates from an increasingly incompatible-element-enriched residual impact melt. This interpretation is consistent with the presence of P-rich glassy melt inclusions trapped in olivine quench crystals, visible in EDS maps (Fig. 5B).

Despite an overall trend, the scatter in P content with olivine Fe-Mg composition across all compositions suggests that P systematics in the shock-melt were highly heterogeneous throughout the crystallization sequence (Fig. 6). This heterogeneity is potentially consistent both with the slow diffusion of P in mafic silicate melt (Watson et al., 2015) and the seemingly non-uniform distribution of P-rich olivine throughout the investigated sample. We infer that the spatial distribution of P-rich olivine in shock-melt lithology derives from the melting of pre-existing phosphates, potentially incomplete homogenization of P through the melt, and finally local P accumulation as an incompatible element during progressive quench crystallization. Future studies may be able to utilize P systematics in olivine as a crystallisation speedometer for quench-cooled shock-melts.

Overall, there is no evidence that any of the phosphates in Chelyabinsk are the products of impact mobilization. Melting did affect phosphates in the shock-melt lithology; however, here, P has been trapped as a minor element in silicate phases (principally olivine) upon cooling and crystallization. The textural evidence we present requires that all preserved phosphate grains existed prior to impact but responded differently to metamorphism depending on lithology-scale conditions and the propagation of shock waves through local phase assemblages. Building on this conclusion, we now highlight textural evidence for the important role played by plagioclase mobilization during impact metamorphism.

## Shock effects on P-bearing minerals

### *Modification of phosphate phase associations*

We explain the anomalous phase associations of Chelyabinsk phosphates using a novel mechanical shock-modification model, which relies upon selective plagioclase melting and its subsequent mobilization towards olivine, pyroxene, and phosphate. We propose that phosphates formed initially during thermal metamorphism with no preferential phase association (as described by Lewis and Jones, 2016). Phosphates in Chelyabinsk then had these initial growth-related phase associations modified during the impact melting and mobilization of easily shock-melted phases such as plagioclase and metals/sulfides (Rubin, 2003a,b). We further propose that, owing to their intergrown nature, phosphate-olivine grain boundaries were mechanically more resilient to plagioclase-melt propagation during impact. The common association of light lithology phosphates with olivine and plagioclase, but infrequent association with metal and sulfides (Fig. 3), provides support for this model.

We observe smooth-textured plagioclase proximal to many light lithology phosphates, as linear trails interstitial to olivine and pyroxene (Fig. 2C), and as thin borders around some silicate grains (Fig. 2C). In places, plagioclase has locally entrained numerous very small (~5 μm) inclusions of silicate, metal, and chromite. Correlating the CL images and EBSD maps further indicates that at least some of this plagioclase is poorly crystalline (Figures 7–8). It is also highly unlikely that surface condition is responsible for the observed poor EBSD response, as other phases respond extremely well. In accordance with previous authors (e.g., Kaeter et al., 2018), we interpret these features as indicative of plagioclase being melted upon initial shock, mobilized in the turbulent flow regime behind the shock front, and quenched by rapid cooling to amorphous material (i.e., maskelynite; Ferrière and Brandstätter, 2015). As such, plagioclase-normative melts appear to have preferentially flowed into pre-existing grain boundaries and fractures opened during passage of the initial shock wave (e.g., Fig. 2C).

A 'preferential-melting model' for increasing the probability of finding plagioclase in association with phosphate in the light lithology only operates over a limited range of shock states. For instance, a much greater percentage of dark lithology phosphates display contacts with metal and sulfide phases than in the light lithology (Fig. 3). This is consistent with the fact that metal and sulfide vein networks are more extensively developed in the dark lithology, as these phases also became mobilized during the higher peak and post-shock conditions experienced (Moreau et al., 2018; Morlok et al., 2017). As such, preferential mobilization of plagioclase resulted in general phosphate separation from metal and sulfide in the light lithology (i.e., destroying growth assemblages), but additional mobilization of Fe-Ni metal and sulfide created a more randomly mixed set of phase associations in the dark lithology. In both cases, associations with olivine were widely retained (Fig. 3).

Our observations suggest that plagioclase in Chelyabinsk was extensively mobilized under S4 conditions, in the light lithology, but that this metal/sulfide mobilization was increasingly extensive under the S5 conditions of the dark lithology. An even stronger shock-deformation would cause wholescale melting of both low-melting point and high-melting point phases and, in particular, would consume pre-existing phosphates. In general, the post-shock phase associations should vary depending upon the phases mobilized. Figure 13 schematically illustrates our preferred model for the sequence of phosphate phase associations in the light and dark lithologies of Chelyabinsk, in the context of the overall evolution of phosphorus-bearing mineralogy in ordinary chondrites.

*Modification of phosphate microtexture*

There is clear evidence for variable phosphate microtexture as a function of different degrees of shock metamorphism, represented by the distinct lithologies in Chelyabinsk. In particular, recrystallization textures in CL and in EBSD are only observed in the dark lithology; although there are sparse SMVs in the light lithology, phosphates proximal to these veins are not recrystallized. The presence of SMVs in the light lithology indicates that SMV-proximal peak temperatures reached > 1950 K for ~70 ms (Ozawa et al., 2014). However, their sparse distribution implies that SMVs would have cooled more quickly in contact with less shocked and less heated matrix in the light compared to the dark lithology, thus, providing insufficient heat for recrystallization of phosphates.

Moreau et al (2018) correlated mesoscale modelling of shock processes to observe textural features in ordinary chondrite samples, finding that the S5-6 transition associated with shock darkening corresponds to > ~1200 K post-shock temperatures in troilite and > ~600 K in olivine. Connecting the shock stage evaluation of Morlok et al (2017) with the results of Moreau et al (2018), peak post-shock temperatures in the dark lithology should have been greater than in the light lithology by a minimum of 200 K, prior to thermal equilibration at much lower temperatures. Given that subdomain formation in a deformed matrix is more efficient at elevated temperatures (Urai et al., 1986), this difference in post-shock heating intensity is a plausible explanation for the contrasting microtextural state of light versus dark lithology phosphates in Chelyabinsk.

Other recent studies have also highlighted the role of shock metamorphism for inducing recrystallization textures in phosphate minerals observed in a wide range of terrestrial (e.g., Cox et al., 2020; Kenny et al., 2020; McGregor et al., 2019) and lunar (Černok et al., 2019) samples. Our findings for the heterogeneously shocked Chelyabinsk breccia suggest that recrystallization textures develop principally in response to prolonged lithology-scale heating during impact-induced strain at the grain-scale. Therefore, we mark out phosphate recrystallization as a sensitive indicator of micro-to-mesoscale *P*-*T*-*t* pathways in the post-impact environment.

It is notable that recrystallization textures are visible in CL (Fig. 9C). Together, the bright internal spots as well as conspicuous bright sub-grain boundaries suggest that the diffusive migration and/or grain boundary enrichment of trace elements during shock processing may be responsible for the specific response displayed (e.g., McGregor et al., 2019; Gros et al., 2016). The EDS images (Fig. 10) do not reveal any such chemical zonation in major elements, but EDS does not resolve variation in the trace elements that typically modulate CL response.

In the scope of this work, we can robustly state that CL imaging provides a way to rapid assess phosphate microtextures prior to EBSD work. As such, our combined CL-EBSD approach allows us to interpret only limited number of high resolution EBSD maps as representative of the diversity of phosphate microtextures present in Chelyabinsk as a whole. Future work should seek to investigate micro-to-nanoscale elemental zonation in shocked phosphates, which may represent an untapped indicator of post-shock cooling rate.

Interpretation of observed CL textures that do not correlate with EBSD textures is challenging. The diminished crystallinity of apatite in contact with plagioclase (Fig. 9A, panel 2-3) may reflect a combination of high transient temperatures achieved by plagioclase-normative melts (Moreau et al., 2018) and grain armoring by high impedance phases (e.g., olivine) in other assemblages (Cox et al., 2020; McGregor et al., 2018). Indeed, we observe patches of small (~5 μm) relict olivine grains in plagioclase (*sensu lato*) throughout both the light and the dark lithologies of Chelyabinsk (Fig. 2C). This is strong evidence for high-$T$ plagioclase-normative melts in the post-shock environment of Chelyabinsk, which should have been capable of preferentially melting and deforming nearby phases (Moreau et al., 2018). The diminished crystallinity of apatite (in both CL and EBSD; Fig. 9) may be one consequence of such proximity. Fracturing is unlikely to be the cause of this poor response, as apatite is regularly fractured and yet most grains we have observed display a strong CL response (Fig. 8A).

Patchy CL texture (Figures 8 and 10) may be superimposed both on dark lithology merrillite recrystallization textures and on apatite grains from light and dark lithologies. This is a novel observation and is difficult to explain. In particular, if both these kinds of CL patchiness share a common origin then it cannot be directly related to $P$-$T$-$t$ pathways during the main shock metamorphic event. No patchy merrillite is visible in the light lithology, but it is observed in apatite in both light and dark lithologies. The patchy texture does show some association with fractures and cross-cutting veins of Fe-Ni-S (Figs. 7 and 10), but our EBSD maps (Fig. 9) do not reveal an obvious structural difference in CL patchy regions of the phosphate minerals themselves.

Ultimately, small scale variation in minor or trace elements around fractures and veins that cross phosphate grain interiors may be responsible for patchy areas, although this will require further spatially-resolved trace-element analysis to confirm. Regardless of the exact casual mechanism, the patchy feature nonetheless helps to infer a two-stage deformation history in Chelyabinsk: (1) an early impact event responsible crystal-plastic deformation of apatite in the light and dark lithologies, merrillite recrystallization in the dark lithology, and P-rich olivine formation in the melt lithology;

(2) a late-stage impact event, potentially related to parent body break-up, that produced overprinting fractures and associated patchy CL textures, in phosphates of both light and dark lithologies (Figures 12 and 13).

We next establish a paragenetic model for phosphorus-olivine associations that spans parent body and impact metamorphism. This model combines our key inferences of metasomatic phosphate formation during parent body metamorphism, the close phosphate association with olivine, a two-stage deformation history for Chelyabinsk phosphates, and dominantly olivine-hosted P in the melt lithology.

**Paragenesis of Phosphorus-Olivine-Assemblages (POAs)**

We find that association between phosphorus and olivine – either as a minor element in quench-crystallized olivine or as discrete phosphate minerals in contact with olivine – is ubiquitous in Chelyabinsk. To systematize these observations, we define a new term for such textures: Phosphorus-Olivine Assemblages (POAs). Earlier, we concluded that phosphate-olivine intergrowth textures in chondrites first develop during parent body thermal metamorphism. The nature of the fluids, exchange reactions, temperature conditions, and timing that mediate olivine replacement by phosphates remain poorly constrained (Jones et al., 2014). Perhaps most puzzling is that olivine, containing few of the major components of phosphates, appears to be the major target material for replacement.

Apatite and merrillite formation during thermal metamorphism must presumably be intrinsically linked to the liberation of phosphorus via oxidation of phosphides such as those still preserved in unequilibrated (low petrologic type) chondrites (Jones et al., 2014). Zanda et al. (1994) identified a trend towards P oxidation during thermal metamorphism and chemical equilibration of primitive chondrites, with siderophile P – initially dissolved as a solid solution in Fe-Ni metal – first bonding with nearby Fe to form Fe-phosphate inclusions.

Subsequent replacement by more stable phosphates involving lithophile cations is an approach towards thermodynamic equilibrium that requires longer-range component migration. Equilibration as tracked by phosphate development can begin during short-term impact-induced migration of fluids but may require heating for up to tens-of-millions of years to reach completion – or, in the case of limited fluid/rock ratio, may never do so (Jones et al., 2016). Indeed, Fe-phosphate inclusions are entirely absent in the type 4 and 5 chondrites, suggesting complete diffusion of P, as well as Cr and Si, out of the metal grains (Zanda et al., 1994). In Chelyabinsk, we observe phosphates in direct association with metal and chromite grains (Fig. 2D; Fig. S1-2). In accordance with previous authors, we therefore suggest that oxidation-driven P diffusion from Fe-Ni grains during thermal metamorphism was a key antecedent of phosphate mineral growth. However, the details of the pathway from reduced P in metal to oxidized P in stable lithophile phosphate minerals is obscure (McCubbin and Jones, 2015; Zanda et al., 1994).

We argue that a short-lived residence of phosphorus in P-rich olivine may constitute a missing link in models for phosphate growth in chondrites. Noting the pervasive association of olivine and phosphorus in equilibrated but unshocked chondrites and in the various lithologies of Chelyabinsk, we define a paragenetic scheme for POAs. Figure 13 illustrates our proposed paragenetic classification model in the context of Chelyabinsk, which is intended for generalized application to POA occurrences in all ordinary chondrites (OCs). We define Type I, II, and III POAs and suggest that they indicate processes of oxidative repartitioning, metasomatism, and post-shock quench crystallization, respectively.

*Type I POAs: products of metal oxidation and metasomatism*

We suggest that the phosphorus-olivine association in chondrites first develops via the formation of P-rich olivine as oxidized phosphorus migrates out of metal grains (Fig. 13A), accompanied by decomposition of Fe-phosphates, during an early stage of thermal metamorphism. We define such metasomatic P-rich olivine as Type I POA (Fig. 13B).

Conspicuous phosphate grains – notably apatites – are found in impact-reheated, low petrologic type carbonaceous chondrites such as DaG 978 (Zhang et al., 2016). Phosphate minerals are otherwise generally absent in such low petrologic type chondrites (see Supplementary Information; Figures S7-8). Instead, DaG 978 contains a unique abundance of low-*T* phosphoran olivine (Li et al., 2017), quite different from the high-*T* P-rich olivine grown from shock melt seen in Chelyabinsk, as well as incompletely developed phosphate minerals. This lends credence to the suggestion that low-energy or distal impact events can play a role in generating stable phosphate phases such as apatite and merrillite by helping to mobilize metasomatic fluids (Treiman et al., 2014; Walton and Anand, 2016).

That low-*T* phosphoran olivine, as seen in DaG 978, should serve as the initial host of P migrating out of metal (Li et al., 2017) appears counter-intuitive, given that under igneous conditions $P^{5+}$ is incompatible in the olivine crystal structure (Shea et al., 2019; Milman-Barris et al., 2008). However, the incorporation of significant amounts of P into olivine has now been observed in a number of disparate settings – mantle xenoliths (e.g., Agrell et al., 1998), pallasites, iron meteorites (e.g., Brunet and Laporte, 1998), quenched shock-melts (Walton et al., 2012), and rapidly crystallized igneous rocks (Milman-Barris et al., 2008). Phosphorus may be incorporated into the olivine crystal structure during disequilibrium igneous processes (e.g., the quenching of a P-rich residual melt; Boesenberg and Hewins, 2010). Phosphoran olivine may also develop in the absence of silicate melt upon oxidation of previously metal-hosted P, e.g., in pallasite and iron meteorites (Van Roosbroek et al., 2017; McKibbin et al., 2016). The same oxidative pathway should be directly applicable to migration from Fe-Ni grains in chondrites (Zanda et al., 1994), and this model predicts the initial formation of an early generation of P-rich to phosphoran olivine during the thermal metamorphism of chondrites. These P-rich silicates would then appear to be reasonable targets for phosphate growth via replacement reactions.

Although the formation of phosphate associated with P-rich olivine in DaG 978 is clearly impact-mediated, the initial oxidation of P and its incorporation in olivine was the result of metasomatism affecting an unequilibrated chondrite. Similar steps likely took place during the thermal equilibration of other chondrites. Our scenario, as in previous models of chondrite oxidation during metamorphism, envisions an oxidizing water-rich mobile phase generated from the breakdown of short-lived ices (e.g., McSween and Labotka, 1993). Interaction with the water-rich fluid partially oxidizes chondritic metal nodules, triggering the repartitioning of $P^{5+}$ (and Cr, Al, etc.) into nearby olivine grains. The resulting Type I POA, a P-rich or phosphoran olivine without distinct phosphate phases and in close proximity to metal grains, is often short-lived but may represent the precursor to later development of phosphate minerals.

*Type II POAs: products of chemical equilibration*

Type II POAs consist of phosphate grains intergrown with olivine (although e.g., pyroxene-derived analogs may also occur – Jones et al., 2016). Type II POAs dominate in the light and dark lithologies of Chelyabinsk. An outstanding question is why olivine, despite its compositional difference to phosphate phases, appears to show such commonplace and mechanically resilient intergrowth with phosphates in Chelyabinsk. Our results offer a possible explanation for this fact in the form of a novel paragenetic framework: specifically, that a number of plausible processes could result in the direct replacement of P-rich silicates by calcium-bearing phosphates.

We propose that Type II POA textures (Fig. 13C) stem from the interaction between Ca-bearing fluids generated during feldspar albitization and the short-lived P-rich olivine paragenesis of Type I POAs. Type I POAs are rarely preserved, being observed so far only in chondrites that experienced short-lived thermal metamorphism and incomplete chemical equilibration (Zhang et al., 2016). These conditions coincide in low petrologic type chondrites that experienced modest shock reheating, such as DaG 978. The Type II POA class is then preserved in the more abundant high petrologic type OCs.

Lewis and Jones (2016) identify a trend from variably Ca-bearing feldspar in low petrologic type OCs through to homogeneously albitic compositions in OC types 5 and 6. Mass balance demands that Ca lost from feldspar is either exported with migrating fluids or locally redistributed to other phases during thermal metamorphism. Important mineral hosts for Ca in OCs include pyroxene, apatite, and merrillite. Notably, clear textural evidence for Ca-bearing pyroxene formation at the expense of olivine is present in Chelyabinsk (Fig. S4). Building on the stoichiometric reaction sequences proposed by Lewis and Jones (2016), we suggest that chondrule and matrix mesostasis, as well as feldspar in more thermally mature cases, acted as Ca-sources to promote secondary mineral formation at the expense of olivine.

Feldspar in type 2-4 OCs has a rough average composition $[Na_{0.4}Ca_{0.6}](Al_{1.6}Si_{2.4}O_8)$, whereas feldspar grains in type 5-6 OCs have compositions that cluster at around $[Na_{0.9}Ca_{0.1}](Al_{1.1}Si_{2.9}O_8)$ (Lewis and Jones, 2016). Given total modal feldspar abundances of 10 % in chondrites, and assuming a density ratio of 0.8 for feldspar in Chelyabinsk relative to the remainder of the mineral assemblage (Reddy et al., 2014), CaO loss from feldspar during thermal metamorphism to type 5-6 equates to a mass loss of 0.8 % as CaO. This would represent > 30 % of the total Ca budget in Chelyabinsk (Righter et al., 2015). This amount of available Ca can easily account for both phosphate growth (< 10 % of total Ca – conservatively estimated based on stoichiometry of merrillite and apatite and a P content of ~0.3 wt%; Righter et al., 2015) and likely also the development of high-Ca pyroxene. Both phosphate and high-Ca pyroxene are observed as olivine-replacing phases in equilibrated chondrites such as Chelyabinsk (Figs. 2A and S4).

Sodium for merrillite formation would be easily provided by locally equilibrating silicates within mesostasis. Magnesium can be derived from olivine itself. Lewis and Jones (2016) noted that apatite growth in OCs appears to be complete by type 4, whereas merrillite growth is a continuous process throughout equilibration to type 5-6. Overall, our model scenario highlights that progressive thermal equilibration plausibly provides sufficient Ca, P, and volatiles (and later Na and Mg) to generate the observed modal abundances of phosphate phases in type 2–6 OCs.

We also include in Type II the assemblages like those in Chelyabinsk, where two-stage shock metamorphism has modified POA phase associations and microtextures (Fig. 13D-E). Type II POAs record information on the *P*-*T*-*t* pathways experienced by chondrite parent bodies, both through parent body thermal and impact-induced metamorphism.

*Type III POAs: products of shock-melt quench crystallization*

We have identified a chemically and textural distinct generation of P-rich quench olivine in Chelyabinsk, which are generally enriched in incompatible elements and have a high density of melt inclusions. We define P-rich olivine grains with a quench crystallization origin as Type III POAs (Fig. 13D). Type III POAs develop during the quench crystallization of shock-melt, where P incompatibility in growing phases leads to P-enriched residual melt. The mechanism of P substitution into Chelyabinsk quench olivine can be inferred from correlated deviations from nominal olivine stoichiometry.

Our results suggest P, alongside Cr and Al, exchanges for dominantly divalent cations, yielding charge vacancies in the olivine crystal structure (Fig. 7). This is in contrast to the Si deficiency often displayed by other P-rich olivine occurrences e.g., Shea et a., 2019; Li et al., 2017. Abundances of Al, Cr, and P are initially linearly positively correlated in zoned olivines; olivines resident in a magma chamber experience progressive diffusion and homogenization of Al and Cr, whereas P homogenization is less efficient (Bosenberg et al., 2004). The positive correlation of Al + Cr with P observed in the Chelyabinsk quench olivine population is consistent with similar observations of rapidly grown olivines with short residence times at high temperature (Milman-Barris et al., 2008). These well-preserved correlations provide further evidence that all of the high-T shock features found in Chelyabinsk were produced simultaneously, with minimal subsequent disturbance.

The textures and chemistry of Type III POAs therefore plausibly represent an as-yet-untapped archive of cooling rate history in quenched shock-melts. We predict that Type III POAs should be abundant in chondrites that experienced high-degree melting, rapid crystallization, and subsequent quiescent thermal/impact histories.

## CONCLUSIONS

We have performed a detailed survey of P-bearing mineral types in the Chelyabinsk impact breccia, including analyses of their abundances, chemical compositions, phase associations, and microtextures. Our textural observations reveal important new aspects of the long-term paragenetic history of P-bearing phases in chondritic asteroids. Correlations between textures imaged in EBSD and CL reveal the latter technique to be a useful tool for rapidly assessing phosphate microtexture. By combining textural and compiled literature evidence, we have revealed the following history of P-bearing phases in Chelyabinsk:

- Initial partitioning of P into Fe-Ni metal grains during chondrite formation
- Low-T formation of P-rich olivine during early parent body metamorphism
- Formation of phosphates via replacement of P-rich olivine during later parent body metamorphism
- Formation of the three lithologies of the Chelyabinsk breccia during a major impact. All melt-related textures in Chelyabinsk are restricted to this single early event.
- All phosphates in light lithology become mildly deformed, merrillites alone in dark lithology become recrystallized. Phosphates are totally melted in melt-lithology; instead, quench crystallization of the melt lithology instead produces P-rich olivine
- Later, phosphate assemblages experience minor disruption by fractures, which also mildly affects their CL response textures. This is likely in response to a second impact.
- Chelyabinsk may have been involved in more collisions than the mere 2 we describe here, but they are not manifest in the textures of the mineral grains we have studied.

Building on our observations, we outline a novel paragenetic classification scheme with which to describe the various forms of Phosphorus-Olivine-Assemblage (POA) found in chondritic asteroids. Type I-II POAs, comprised of metasomatically produced P-rich olivine and intergrown phosphate and olivine, respectively, provide information on pre-impact radiogenic heating, volatile chemistry, and chemical equilibration within parent body asteroids. Type II POAs also have the potential to constrain impact-induced events in asteroidal parent bodies. Finally, Type III POAs, comprised of P-rich olivine formed at high temperatures, correspond to the quenching of chondritic shock-melt to produce spatially, texturally, and chemically heterogeneous P-enriched olivine grains throughout individual meteorite samples.

Important unanswered questions about the setting-specific response of phosphates to shock metamorphism include the specific chemical mechanisms of olivine replacement by phosphate minerals, the boundary conditions that correspond to different shock textures, and the U–Pb systematics of these texturally heterogeneous shocked phosphates. However, these questions should be looked upon as unique opportunities to further constrain complex collisional histories. Future work should seek to leverage the combined temporal and physiochemical records retained by POAs, in order to resolve the timing and extent of shock metamorphism in parent body asteroids.

**Acknowledgments**

We gratefully recognize the support given by Sam Hammond and Dan Topa during EPMA work, Gordon Imlach during EBSD-imaging, and Diane Johnson, Giulio Lampronti, and Iris Buisman for assistance with SEM and CL work. Thanks also go to Colin H. Donaldson, Helen Williams, Sami Mikhail, Darren Mark, and Ryan Ickert for helpful scientific discussions during the course of this work. Thanks to Julia Walter-Roszjár for providing the BSE map of one of the investigated sections. Special thanks go to Richard Greenwood for initial guidance in the early stages of this work back in 2016, as well as for suggesting Chelyabinsk as a sample to work on. We thank Agata Krzesińska and an anonymous reviewer for helpful and constructive review comments, and Christian Koeberl for editorial handling. M.A. thanks the Royal Astronomical Society (RAS) for providing a Paneth Trust for Meteorites Research summer internship through which C.W. initiated this research project. C.W.

acknowledges NERC and UKRI for support through a NERC DTP studentship, grant number NE/L002507/1. M.A. acknowledges support from a UK Science and Technology Facilities Council (STFC) grant (#ST/P000657/1). I.B. greatly acknowledges SYNTHESYS (www.synthesys.info) – a European Union-funded Integrated Activities grant – for support of this research. A.Č. and M.A. acknowledge funding from the European Union's Horizon 2020 research and innovation programme under grant agreement No-704696 RESOLVE. The authors declare no conflict of interest.

**Figures:**

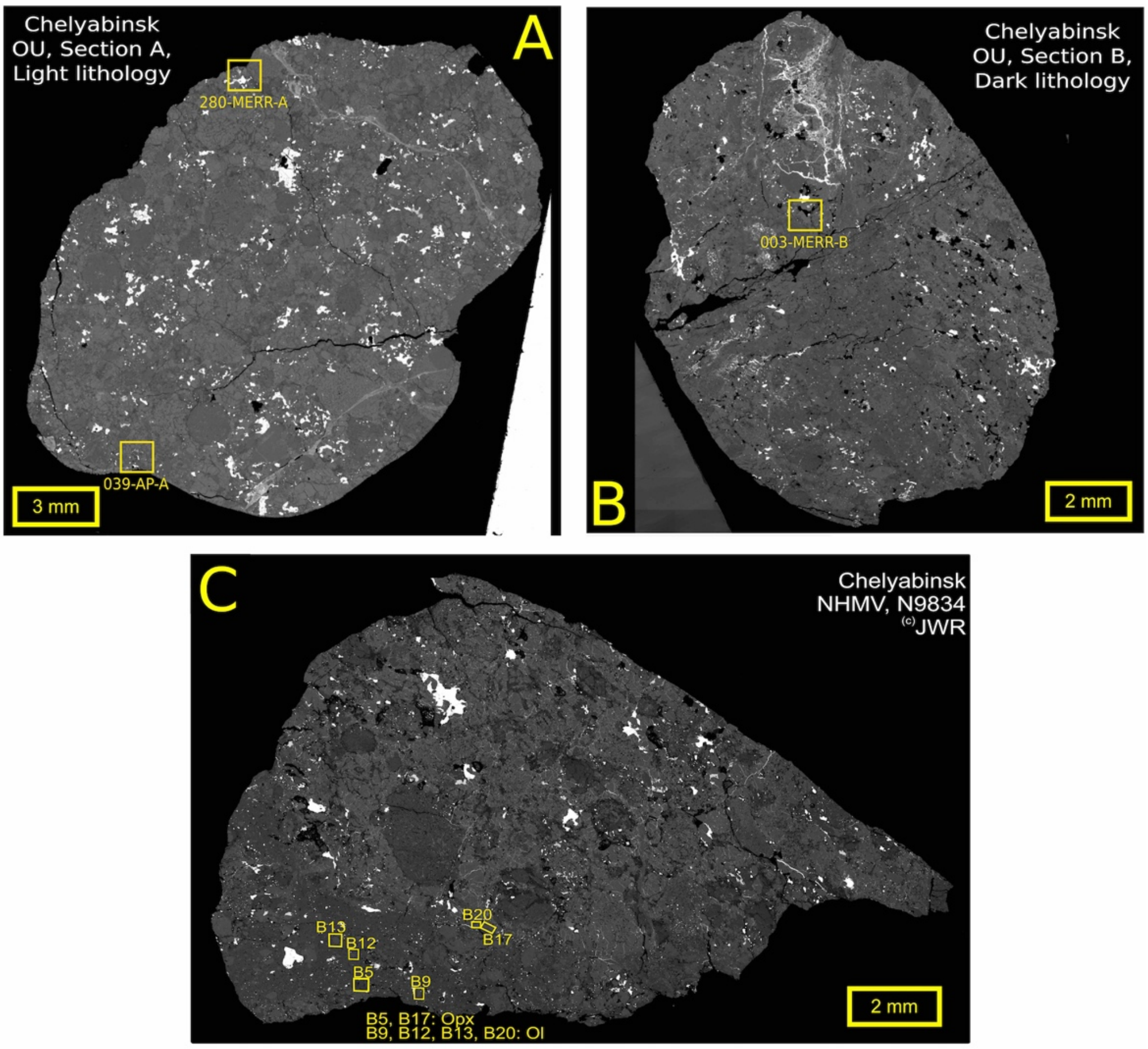

**Figure 1:** Backscattered-Electron maps of Chelyabinsk light (A), dark (B), and shock-melt lithology (lower left-hand side of section N9834 (C), which is otherwise comprised of the dark lithology). Phosphate/P-bearing olivine grains analyzed with EBSD are indicated, though many more were studied with SEM. Phosphate grains are absent in the shock-melt lithology. Together, these three sections span a range of shock stages (S4-6), thus preserving in pristine condition the textural outcome of exceptionally varied pressure-temperature-time pathways throughout the meteorite.

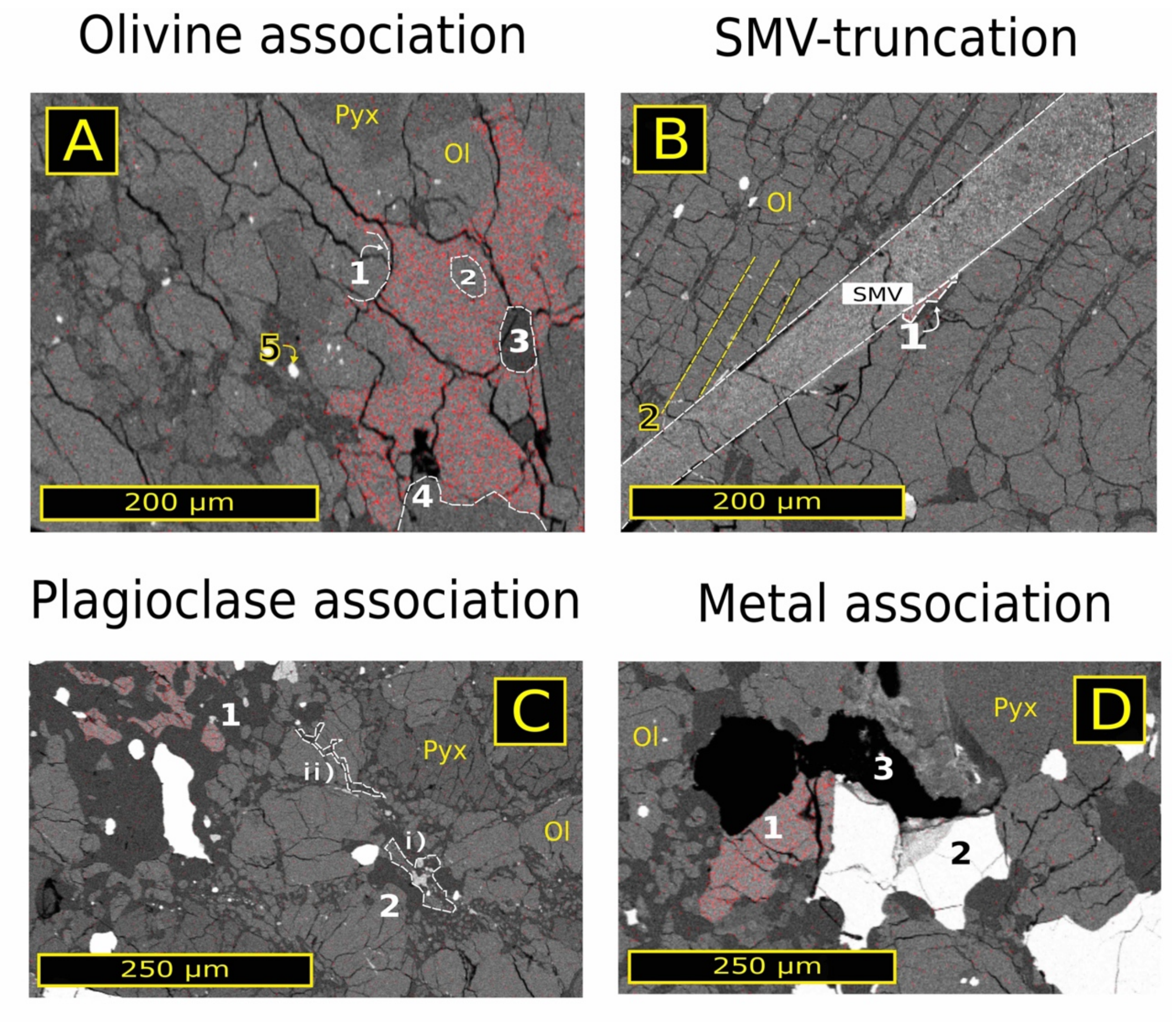

**Figure 2**: Backscattered-Electron images combined with EDS P element intensity maps (in red) showing key phosphate phase associations in the light lithology. (A) Grain 026-MERR-A, showing close association and complex intergrowth texture of merrillite with olivine. Specific textures are numbered as follows: (1) curvilinear olivine/phosphate boundaries, (2) olivine inclusions, (3) a grain of plagioclase, (4) pyroxene, and (5) Fe-Ni metal. (B) Apatite 079-AP-A, showing truncation of apatite (1 – black dashed outline) and barred olivine (2 – yellow dashed outlines) by shock melt vein (SMV; white dashed outline). (C) Apatite 039-AP-A, with textural evidence for plagioclase mobilization. A pool of plagioclase containing large apatite and metal grains (1) is shown connected to other plagioclase pools by a variety of mobilization textures (2), including zones marked by finely interspersed olivine and chromite in plagioclase (i), as well as thin rims around and linear fills of apparent planar deformation features in olivine and pyroxene (ii). (D) Merrillite grain 061-MERR-A (1) closely associated with a metal grain (2) and a void (3). Mineral abbreviations: Olivine = Ol; Pyroxene = Pyx.

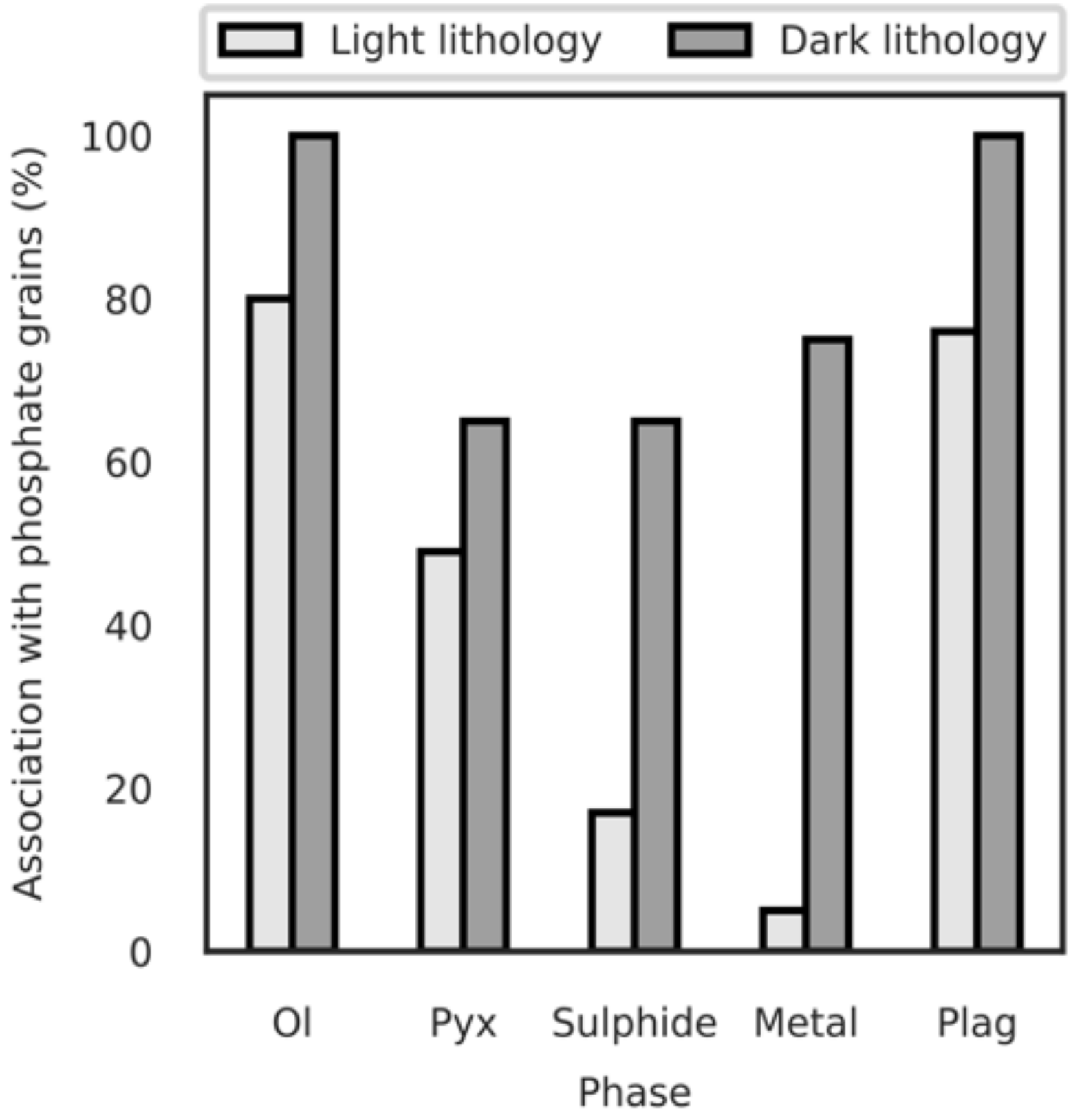

**Figure 3:** Counting statistics for association of phosphate grains with other phases in the light and dark lithologies of Chelyabinsk (n = 96 individual phosphate grains), based on an SEM survey. Mineral abbreviations: Olivine = Ol; Pyroxene = Pyx; Plagioclase = Plag.

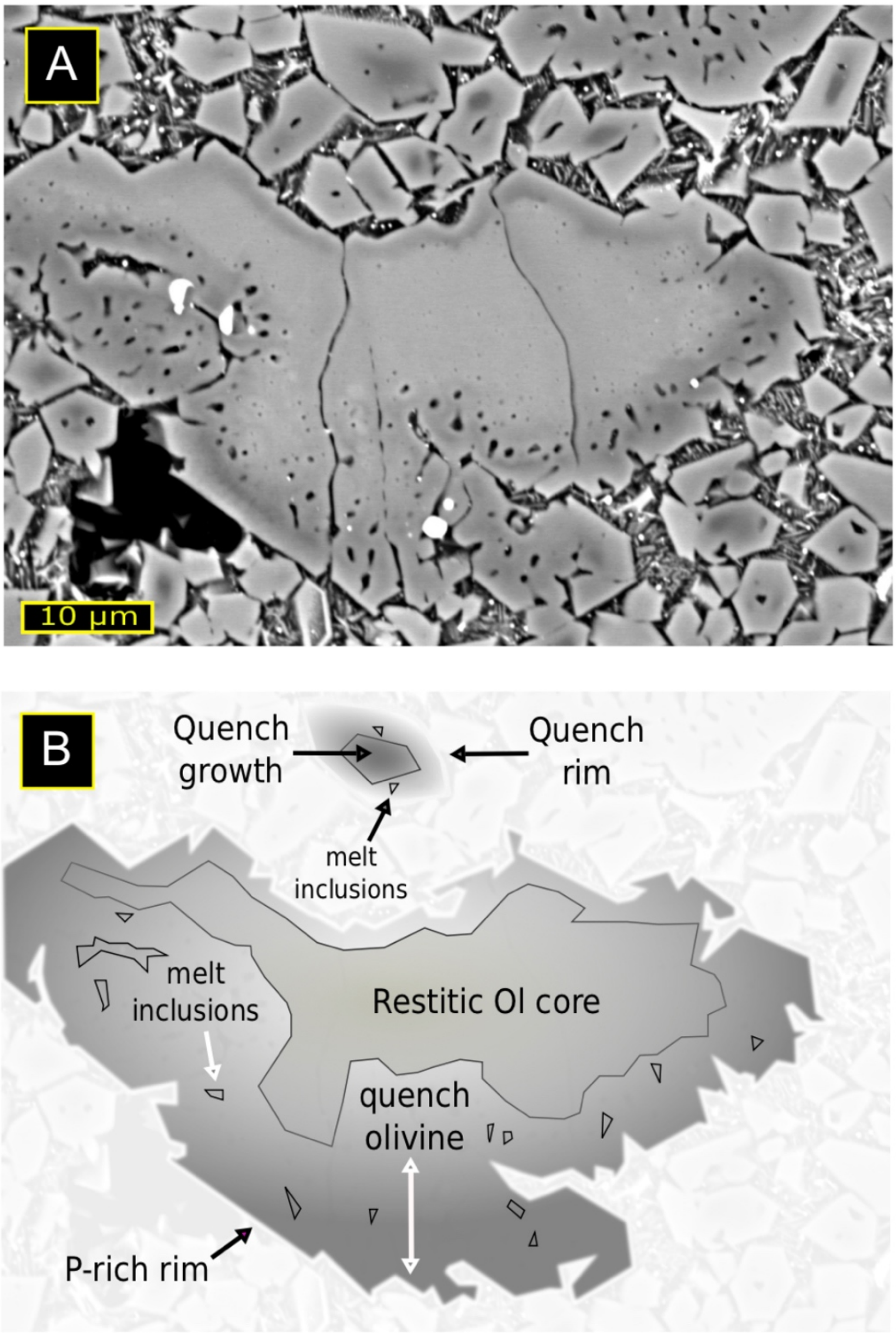

**Figure 4:** (A) Backscattered-Electron (BSE) image of olivine grain B6 and surrounding quenched melt. (B) schematic view of the same olivine grain shown in (A). Two populations of olivine in this lithology are distinguished: partially resorbed relics and quench olivine, found as individual crystallites and rim-overgrowths around partially resorbed relics. Quench olivine is clearly zoned, and is further subdivided into individual crystals and rim-overgrowths that formed around pre-existing olivine crystals. Mineral abbreviations: Olivine = Ol.

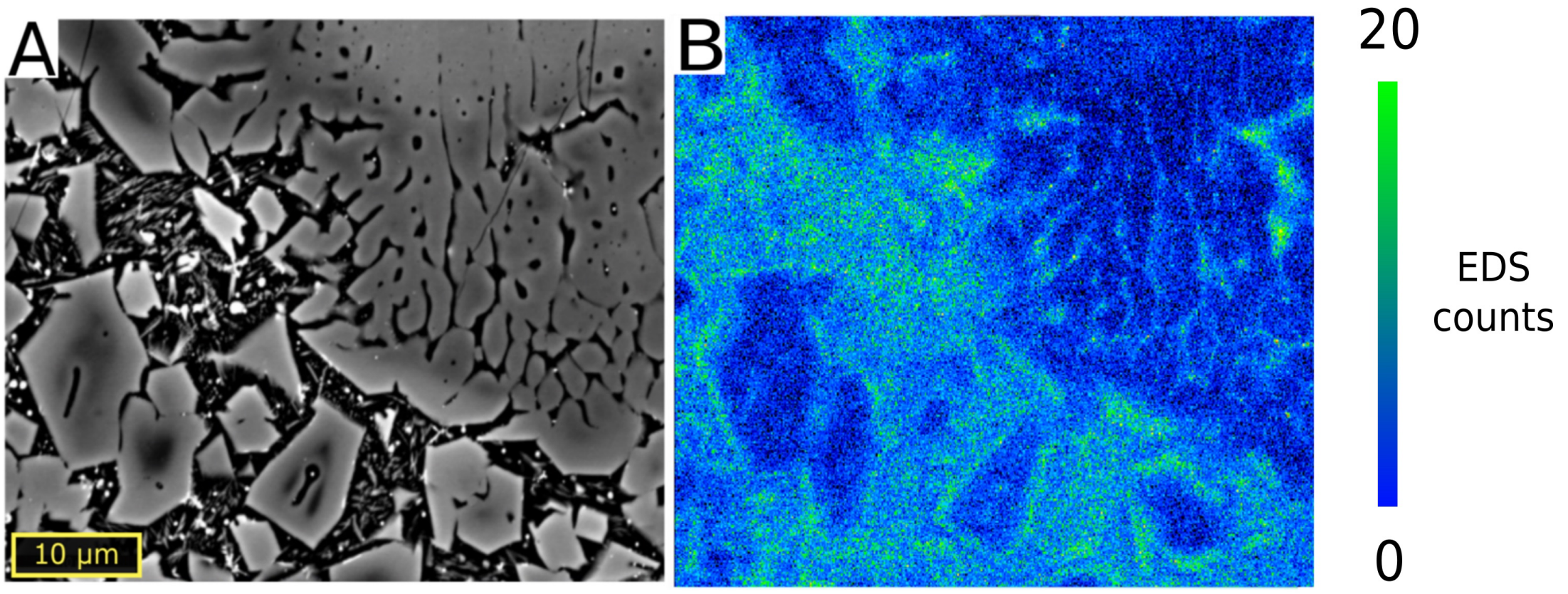

**Figure 5:** Backscattered-Electron image (A) and P EDS intensity map (B) (scale is in counts) from the rim of an olivine grain (B12) occurring in the shock melt lithology. Highest P concentration is observed at grain-edges.

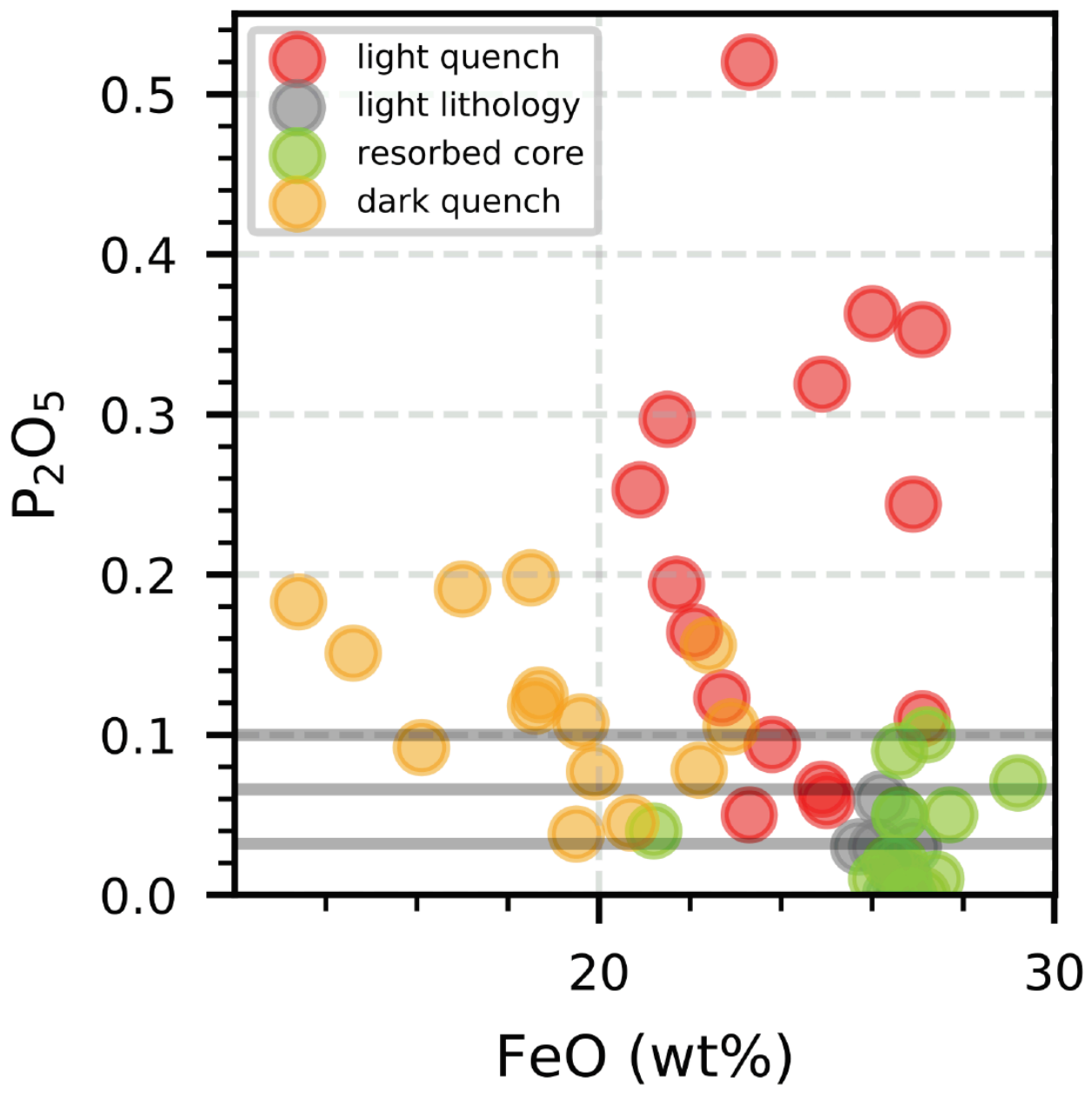

**Figure 6:** FeO versus $P_2O_5$ content of olivine populations from Chelyabinsk. Early (dark) quench olivines are relatively Fe-poor and variably P-poor to moderately P-enriched, compared to pre-shock olivines of both dark lithology resorbed and undisturbed light lithology type. Late (light) quench olivines are variably P-poor to highly P-enriched, with Fe-contents that range from slightly lower-than to equivalent-to the pre-shock populations. Grey lines indicate the 1, 2, and 3 σ detection limit for $P_2O_5$.

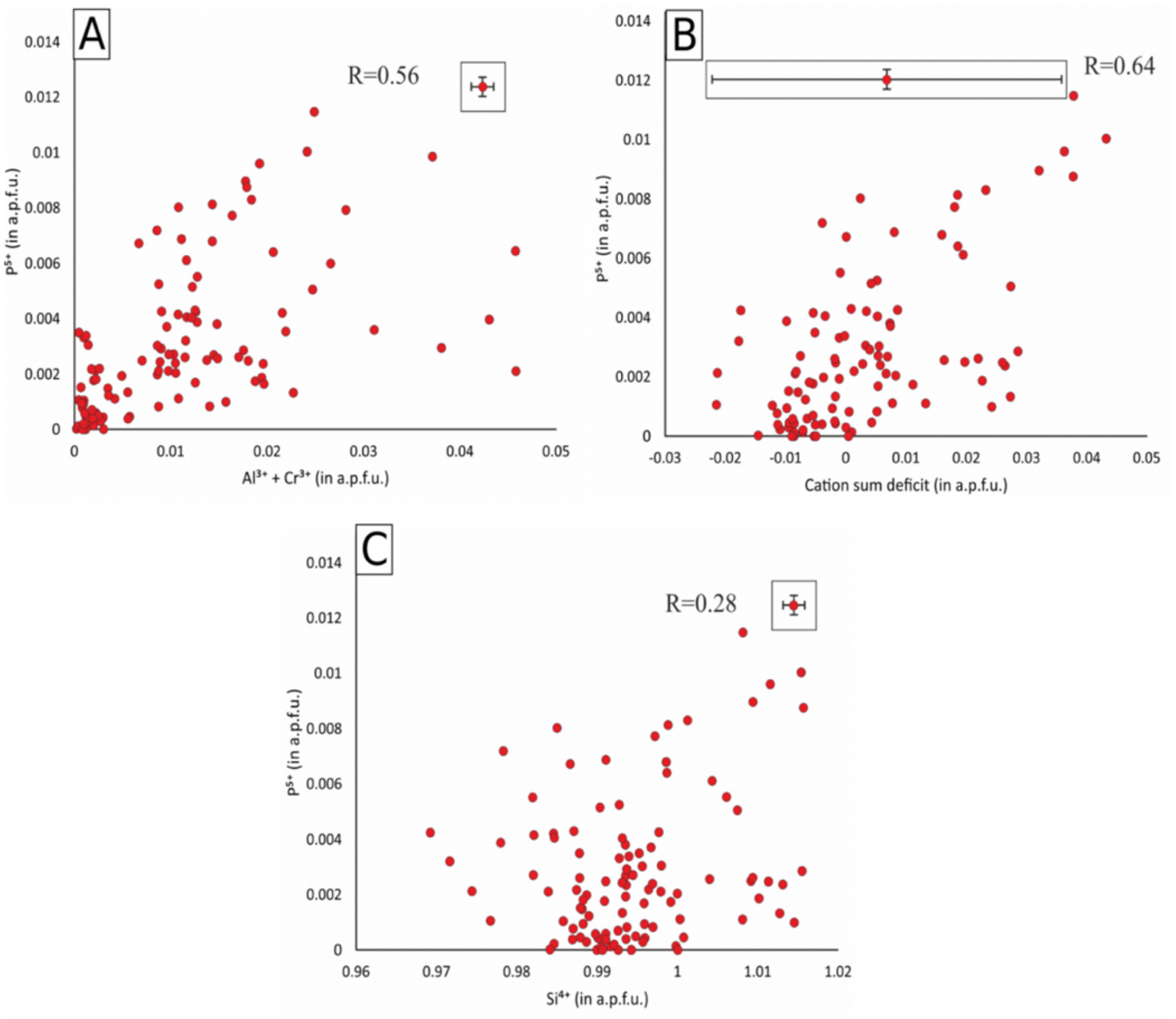

**Figure 7:** Olivine compositional data (as obtained with microprobe) showing correlations or lack-thereof between phosphorus and (A) $Al^{3+} + Cr^{3+}$, (B) cation sum deficit, and (C) $Si^{4+}$.

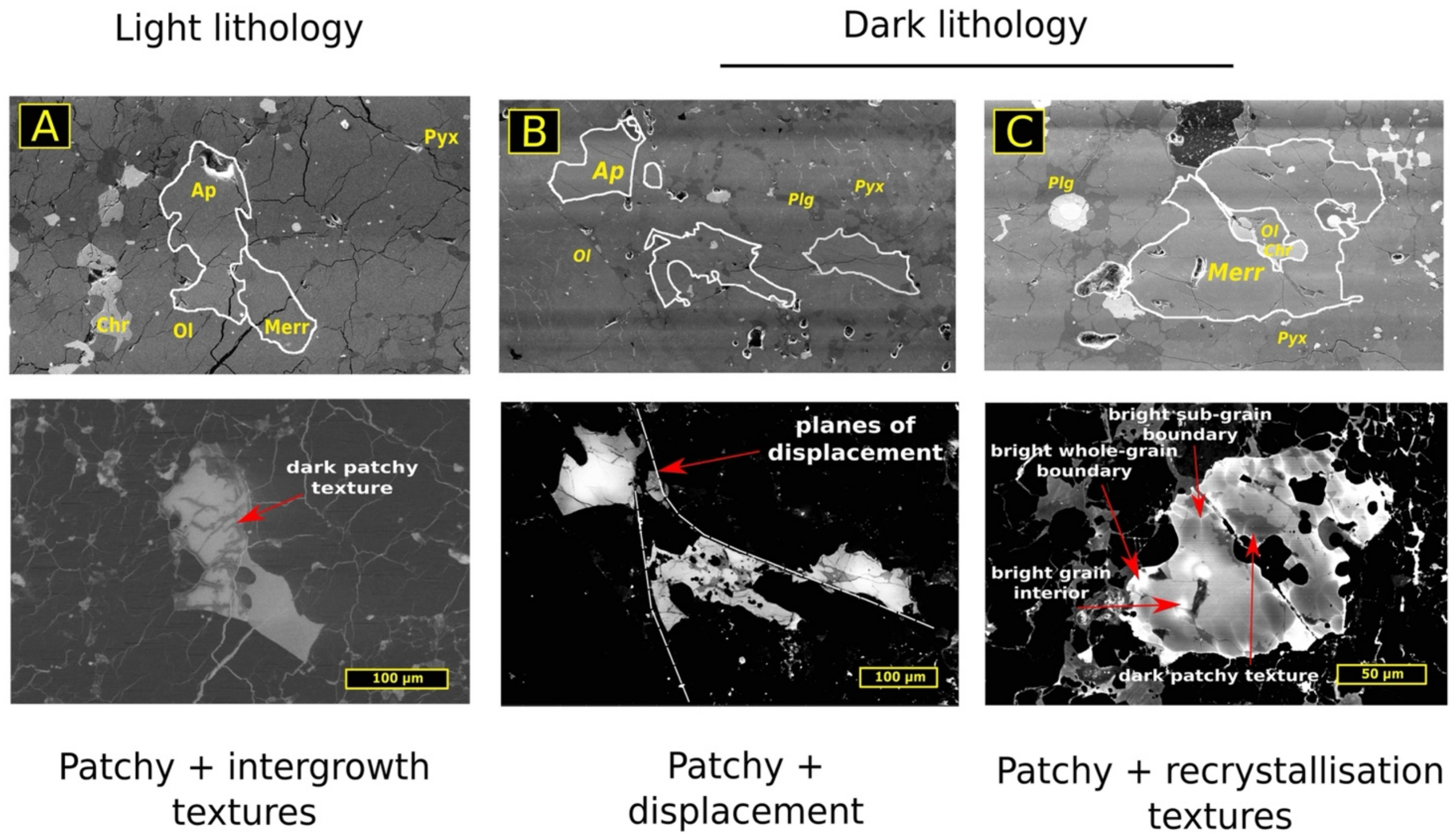

**Figure 8:** Secondary electron (SE) and Cathodoluminescence (CL) images, top and bottom, respectively, showing phosphate textures in Chelyabinsk. (A) 191-MERR/AP-A showing apatite with complex CL response intergrown with a merrillite with contrastingly smooth CL response. (B) 018b-AP-B showing light and dark zones in a patchy texture, as well as micro-fault displacement textures on the grain-scale. (C) 003-MERR-B shows variably dark and bright CL response. Bright regions occur on the whole-grain boundary, some subdomain boundaries, and as internal bright spots. There are also apparently overprinting patchy regions of dark CL-response. Mineral abbreviations: Olivine = Ol; Pyroxene = Pyx; Apatite = Ap; Merrilite = Merr; Plagioclase = Plag; Chromite = Chr.

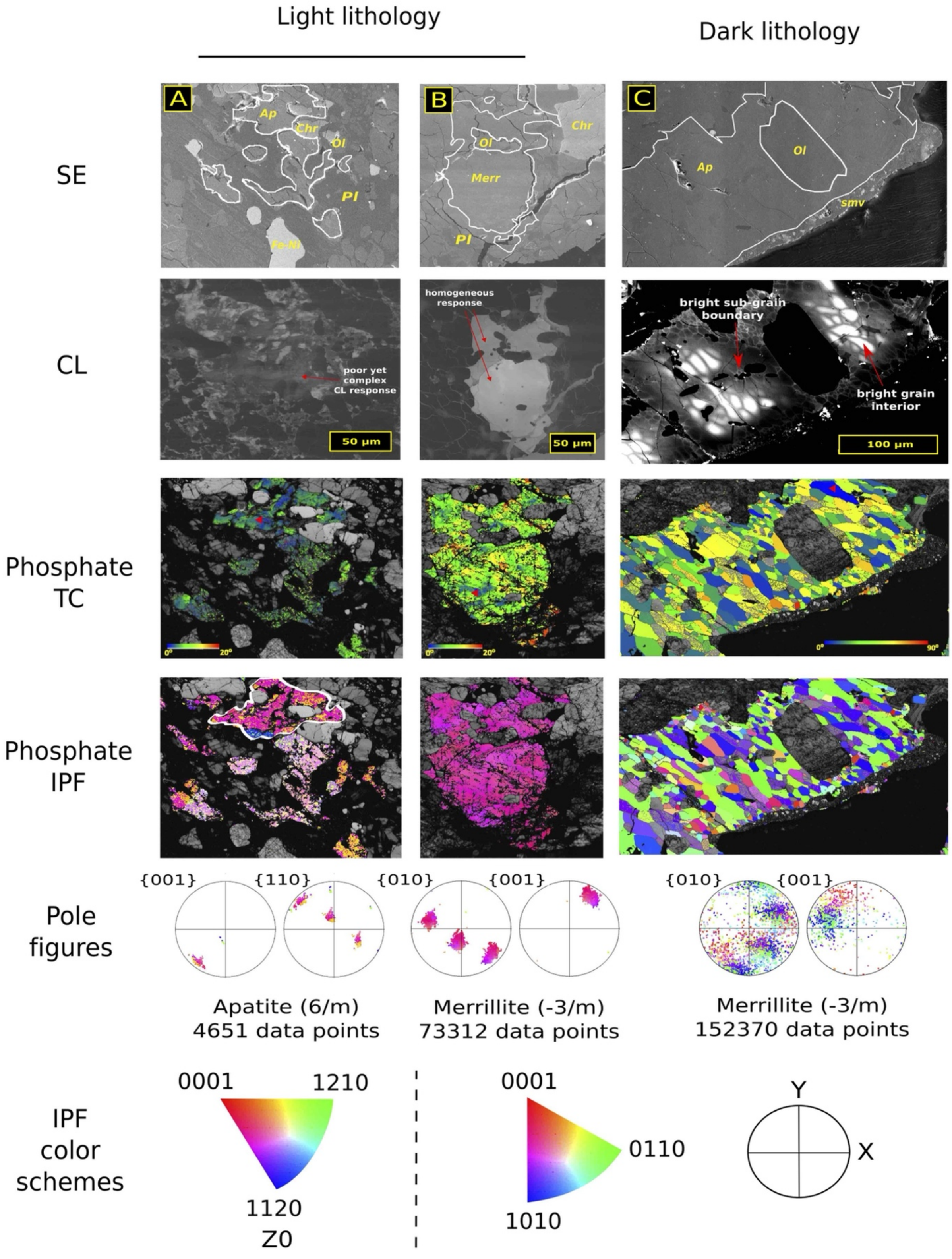

**Figure 9:** Secondary Electron (SE) annotated phase map, Cathodoluminescence (CL) images, Texture Component (TC) maps, pole figures, and Inverse Pole Figure (IPF) colormaps, for selected phosphates. A red or blue triangle indicates the TC reference point in respect to which the internal grain misorientation is measured. All pole figures shown are color-coded after the corresponding IPF map. [Chr = Chromite; Fe-Ni = iron-nickel metal; Ol = Olivine; Pl = plagioclase; SMV = shock melt vein]. (A) 39-AP-A showing complex internal structure of apatite with complex CL response. This grain is surrounded by plagioclase. Several apatite grains showing distinct lattice orientations, according to IPF maps. The pole figure of the selected apatite grain (highlighted in white) shows moderate crystal-plastic deformation. (B) Grain 280-MERR-A shows a bright CL response with minor subdomain heterogeneity as well as plastic deformation with up to 16 degrees misorentiation. The pole figure indicates continuous deformation of the lattice, but no randomly oriented subdomains. (C) Recrystallised merrillite from dark lithology displays a complex internal network of strain-free subdomains (visible in CL; confirmed with EBSD). The pole figure indicates that the orientation of individual subdomains broadly follows the orientation of the parent-grain lattice, but with significant scatter.

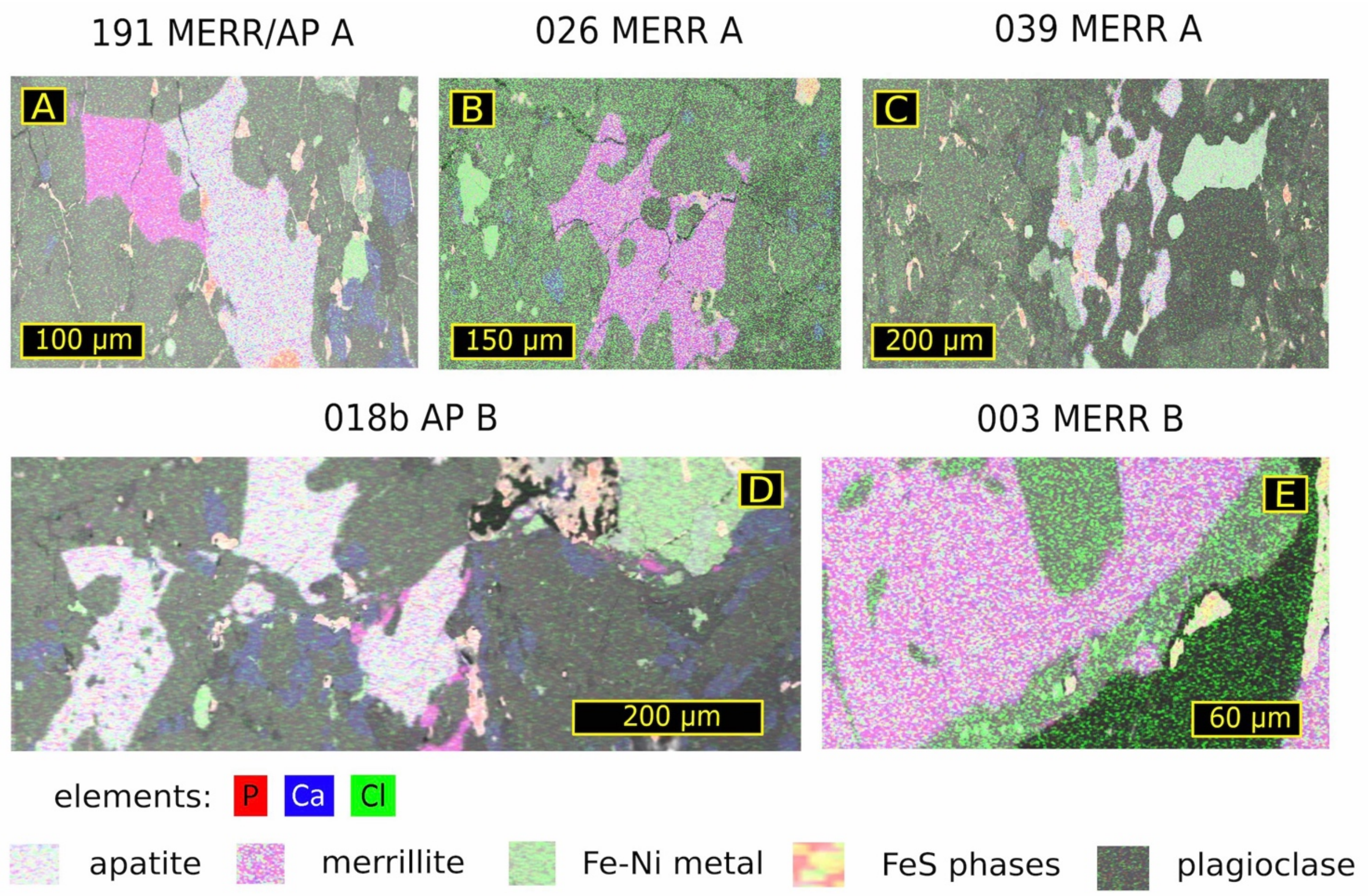

**Figure 10:** Energy dispersive X-ray maps + backscattered electron images of selected phosphate grains from light and dark lithology (sections A and B, respectively). Grains show a homogeneous major element chemical response regardless of their internal microtextural state in CL or EBSD (A) Smooth merrillite and complex apatite grain (in CL). (B) Smooth merrillite grain (in CL). (C) Degraded apatite grain (in CL). (D) Patchy apatite and recrystallized merrillite grain. (E) Recrystallized merrillite grain (in CL and EBSD).

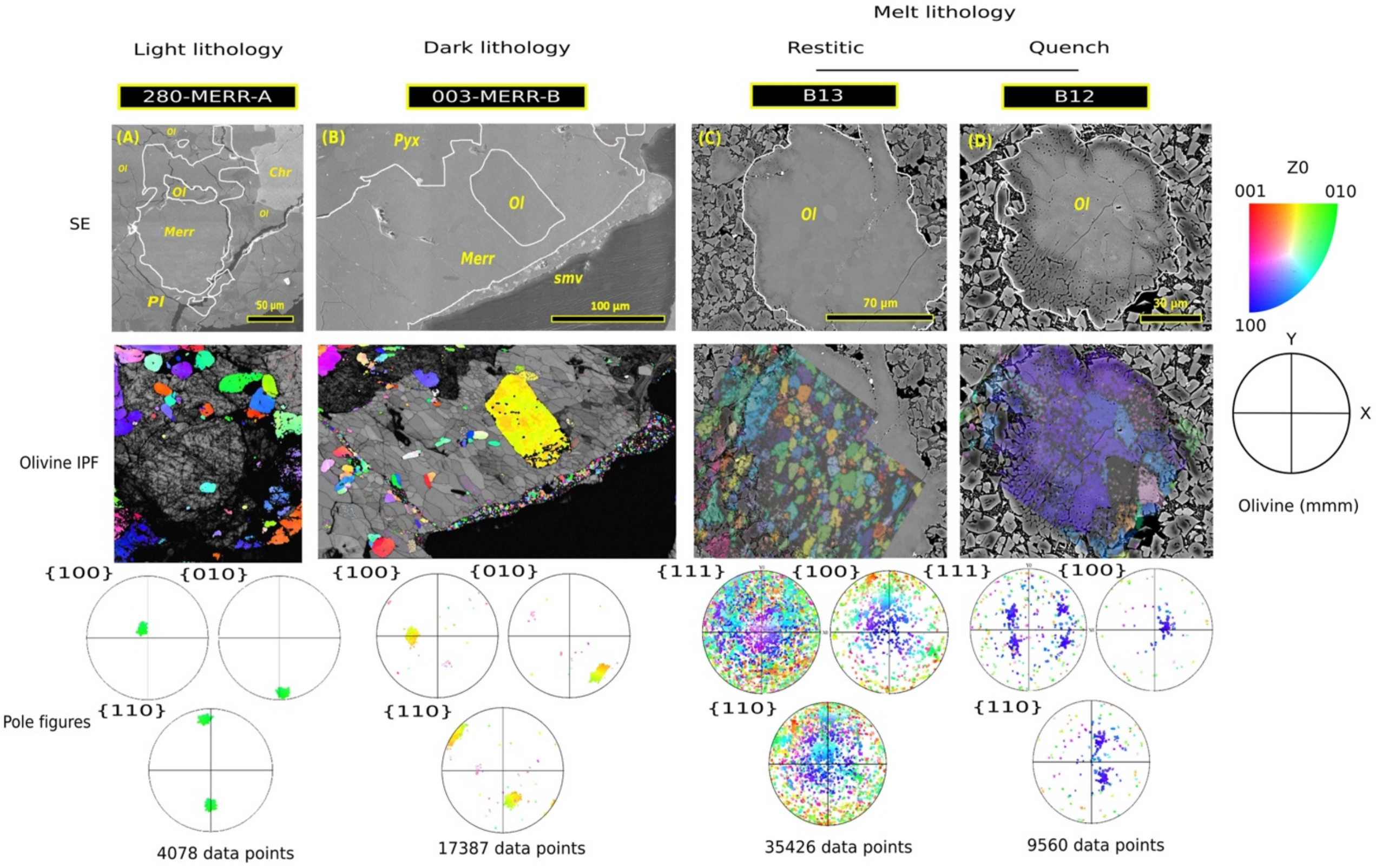

**Figure 11:** (A) Inverse Pole Figure (IPF) map of olivine surrounding and included within MERR-280-A reveal random orientations and minimal deformation of individual grains. (B) IPF map of olivine inclusions in MERR-003-B reveals that individual grains are randomly oriented. The pole figures of a large (outlined) olivine inclusion indicate that the entire grain shows moderate crystal-plastic deformation, but no randomly oriented subdomains that would be indicative of recrystallization. (C) IPF maps and pole figures for olivine grain B13 reveal numerous subdomains, indicative of recrystallization. Data are for the orthorhombic structure of olivine of mmm crystal class. (D) IPF maps and pole figures for olivine grain B12 reveals larger strain-free subdomains, indicative of crystallization from a melt without subsequent deformation. Mineral abbreviations: Olivine = Ol; Pyroxene = Pyx; Apatite = Ap; Merrilite = Merr; Plagioclase = Pl; Chromite = Chr.

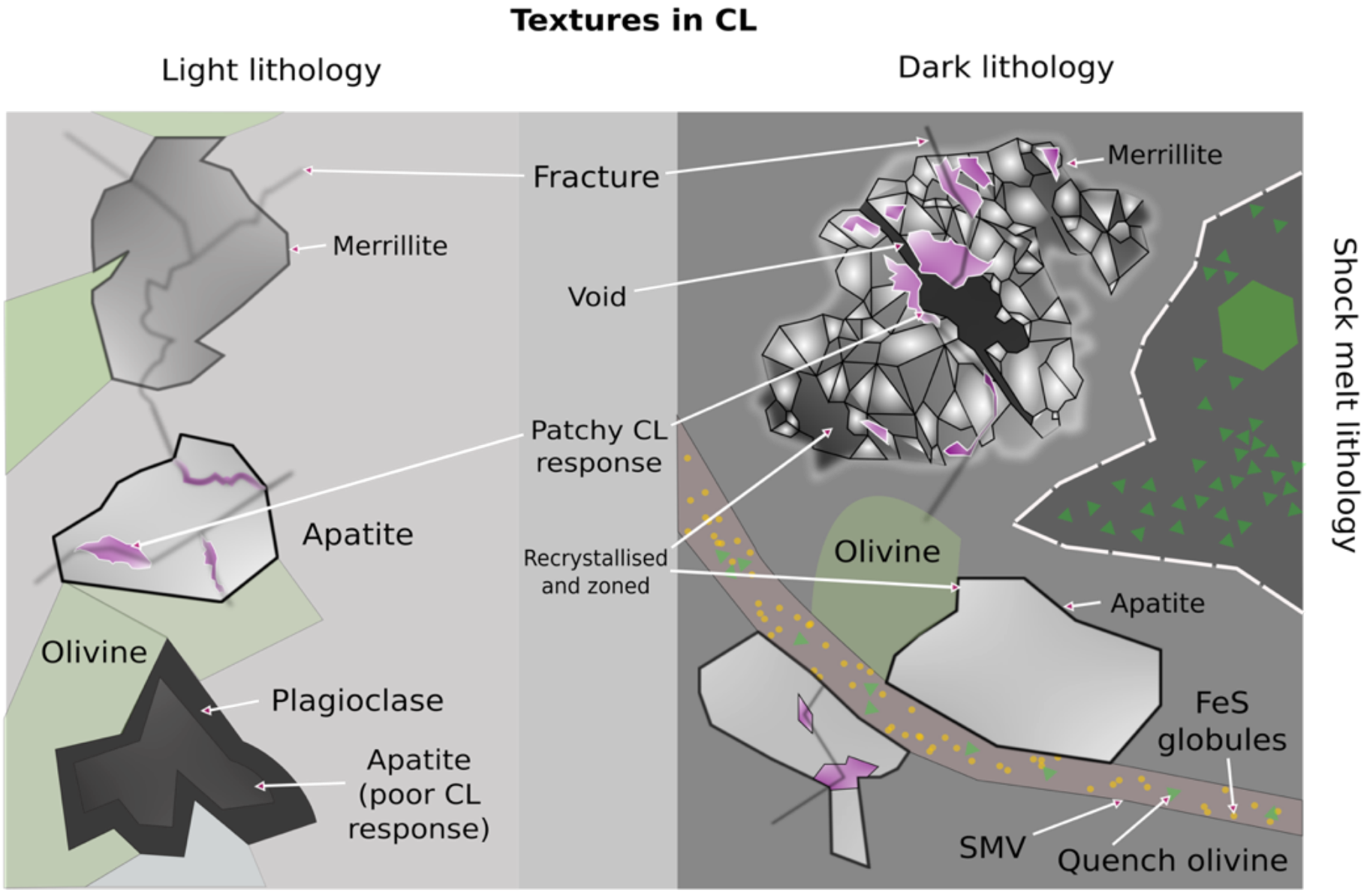

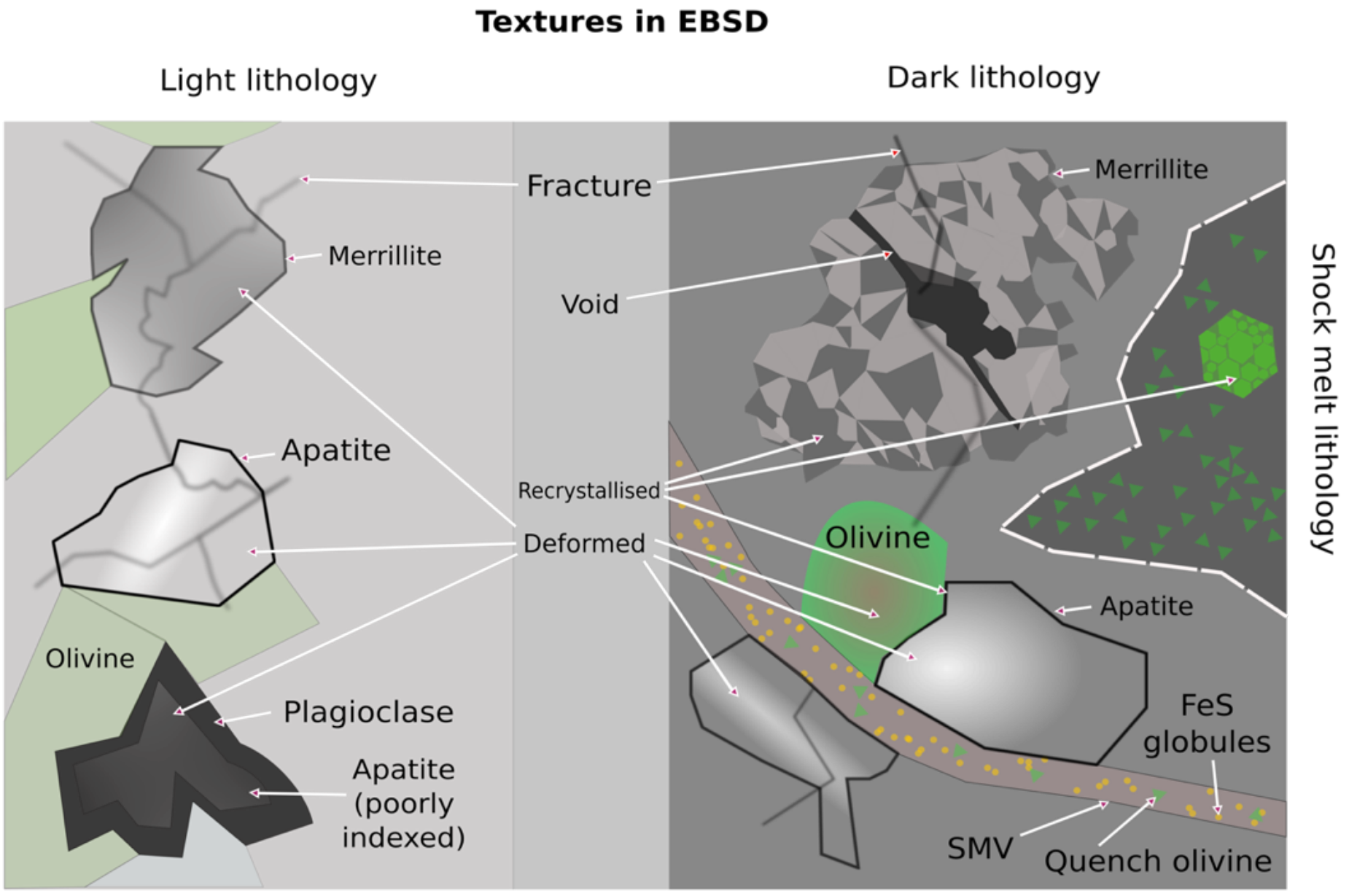

**Figure 12:** Generalized phosphate microtextures observed in Chelyabinsk with (top panel) CL imaging, which consistently reveals the widest diversity of features, and (bottom panel) EBSD maps, which show strained and unstrained (recrystallized) phosphates and olivine grains.

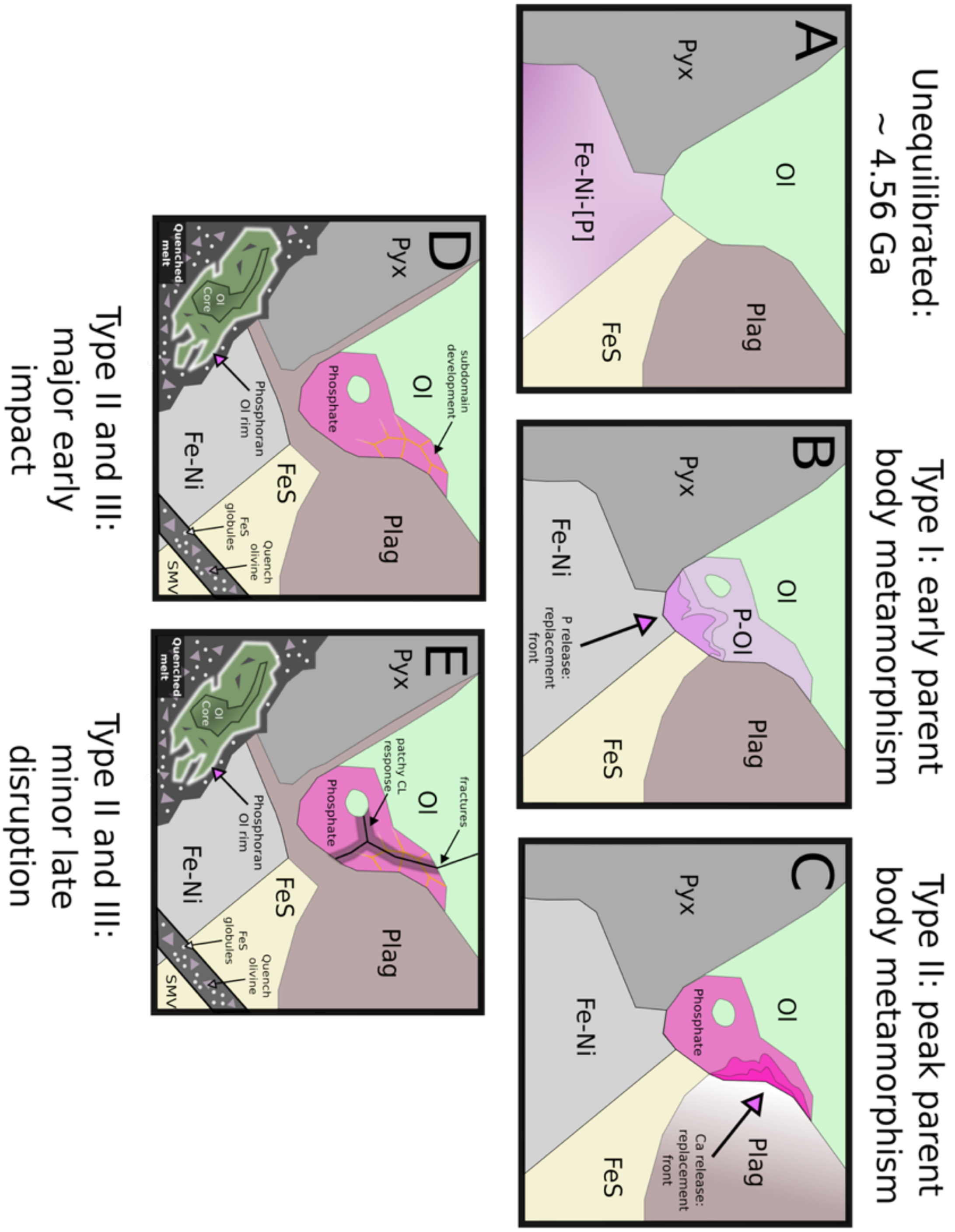

**Figure 13:** Schematic evolution of phosphorous mineralogy in chondrites. (A) ~4.56 Ga – unequilibrated assemblage of minerals, with P bound mainly to metal in a siderophile form. (B) Type I POAs: early stage radiogenic parent body metamorphism, oxidation and release of P from metal, and the consequent formation of P-rich olivine (or, alternatively, during impact induced alteration of low petrologic type material, at any time). (C) Type II POAs: chemical equilibration during further parent body heating. Multistage phosphate growth via olivine-replacement. (D) Shock metamorphism produces new textural features for type II POAs, including subgrain development (in merrillite), and new phase associations (with e.g., mobilized plagioclase). Type III POAs form in shock-molten regions as P-rich rims on restitic recrystallized cores or quench microlites of olivine. (E) A later minor disruption event fractures the mineral assemblages, inducing patchy CL response in phosphates of both lithologies. Mineral abbreviations: Olivine = Ol; Pyroxene = Pyx; Plagioclase = Plag.

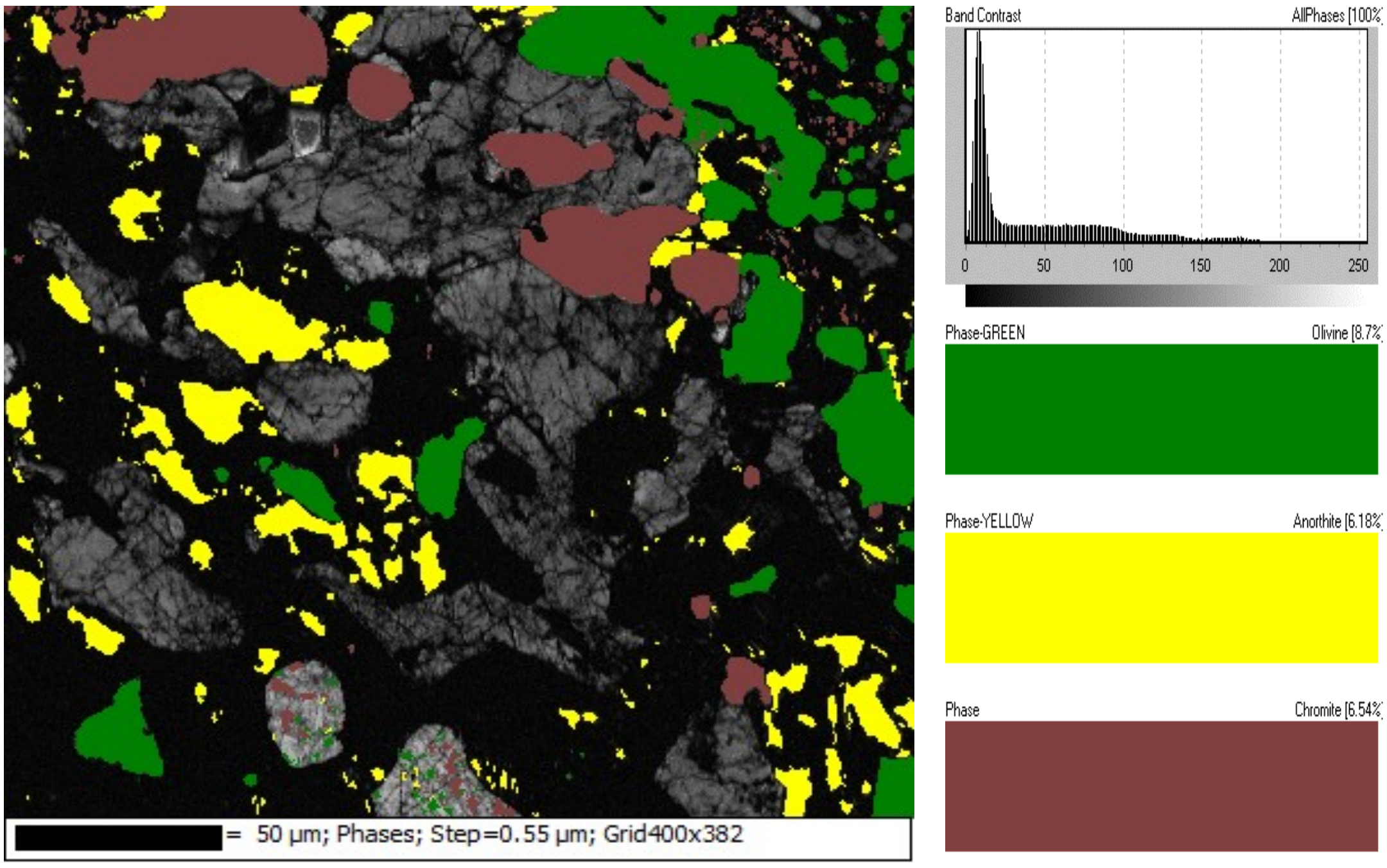

Figure S1: Band Contrast (BC) + EBSD phase map of apatite grain in light lithology (AP-039-A), showing textural evidence for phosphate association with chromite and plagioclase. Plagioclase is largely non-indexable (black). Phosphate is grey. Other phases are color-coded.

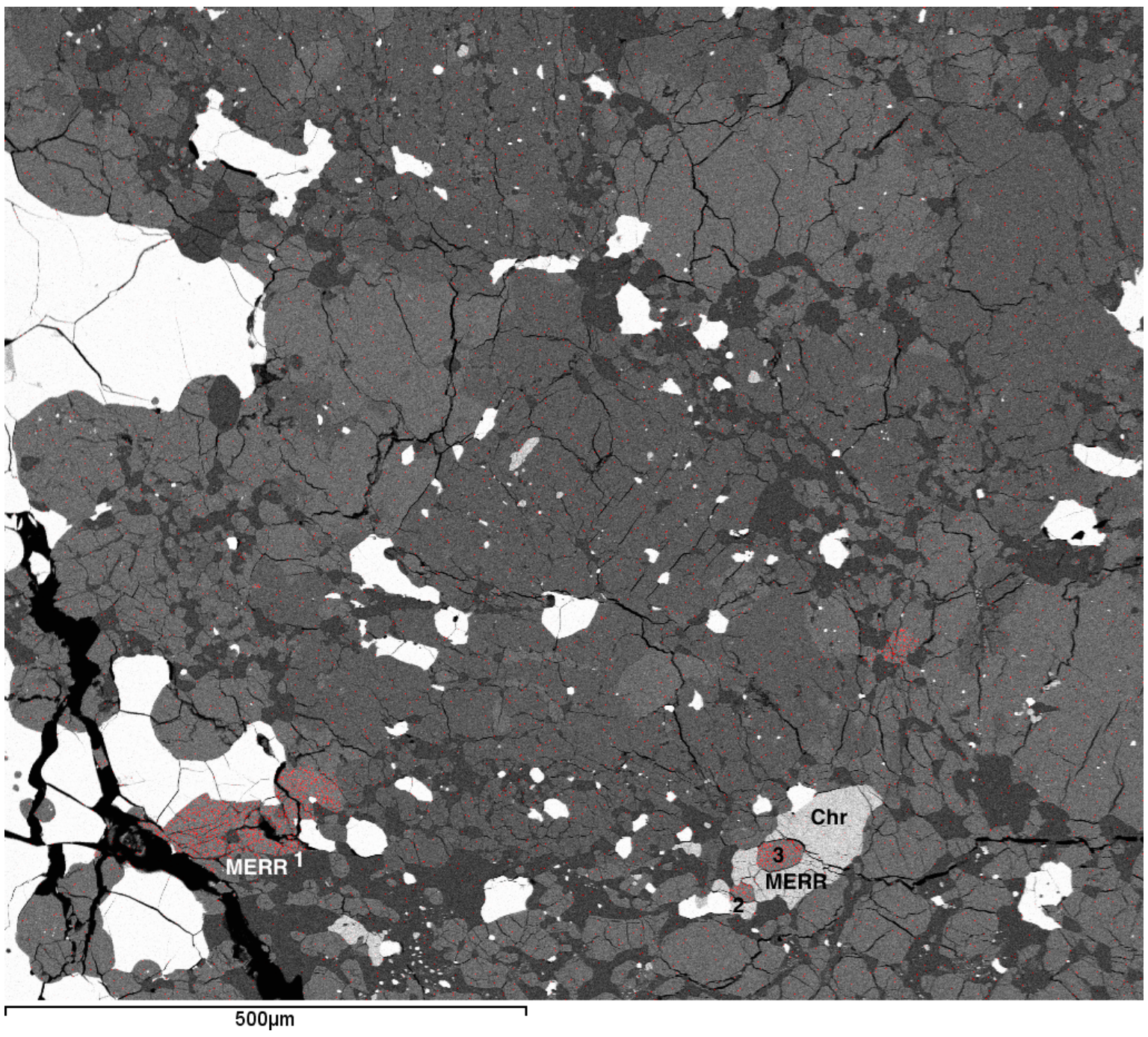

Figure S2: Backscattered electron image 231 of lithology A – textural evidence for phosphate (MERR-2/3) association with chromite. Mineral abbreviations: Merrilite = Merr; Chromite = Chr.

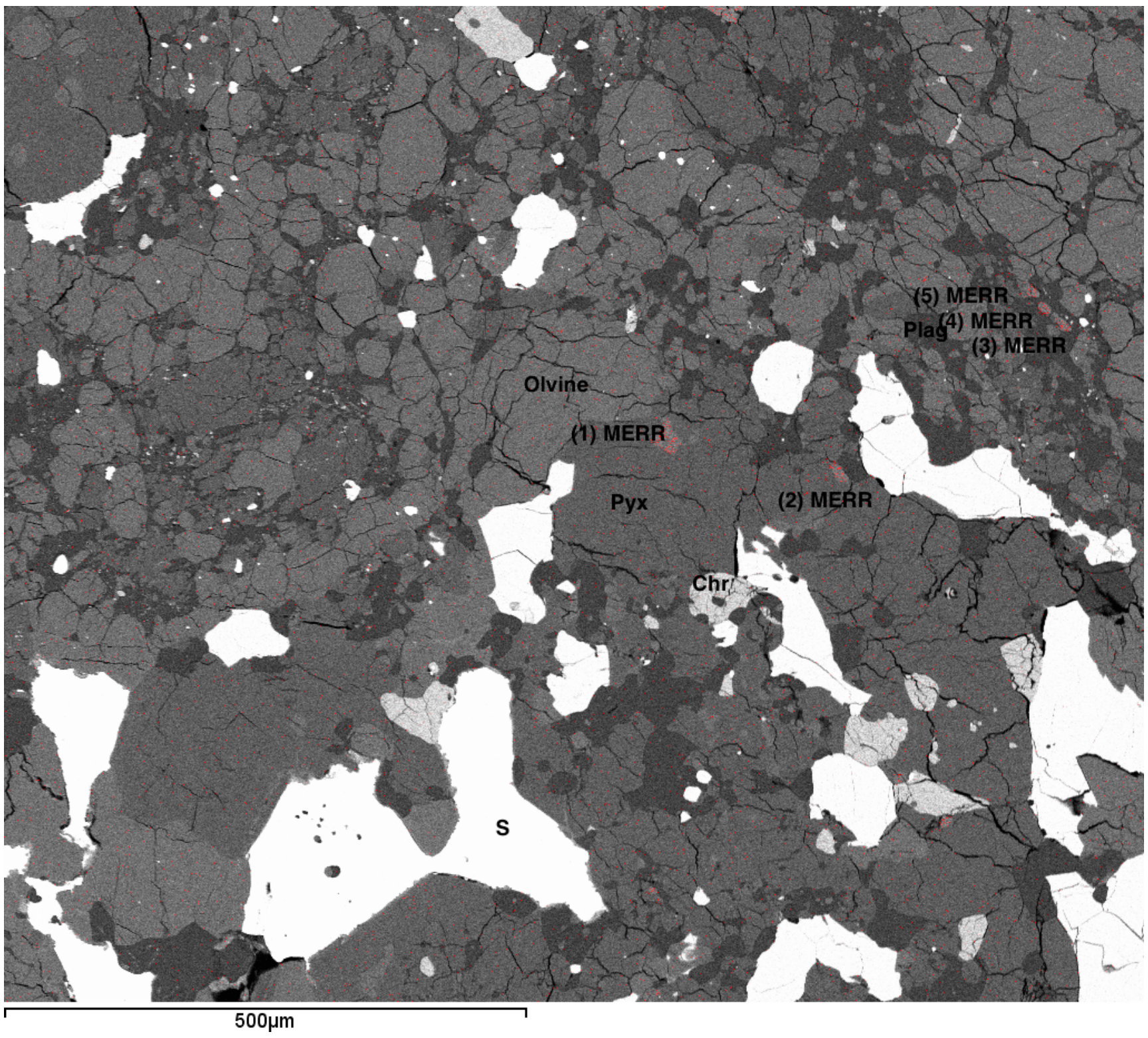

Figure S3: Backscattered electron image P78 of Lithology A, showing phosphate inclusions within pyroxene. Mineral abbreviations: Olivine = Ol; Merrilite = Merr; Plagioclase = Plag; Chromite = Chr.

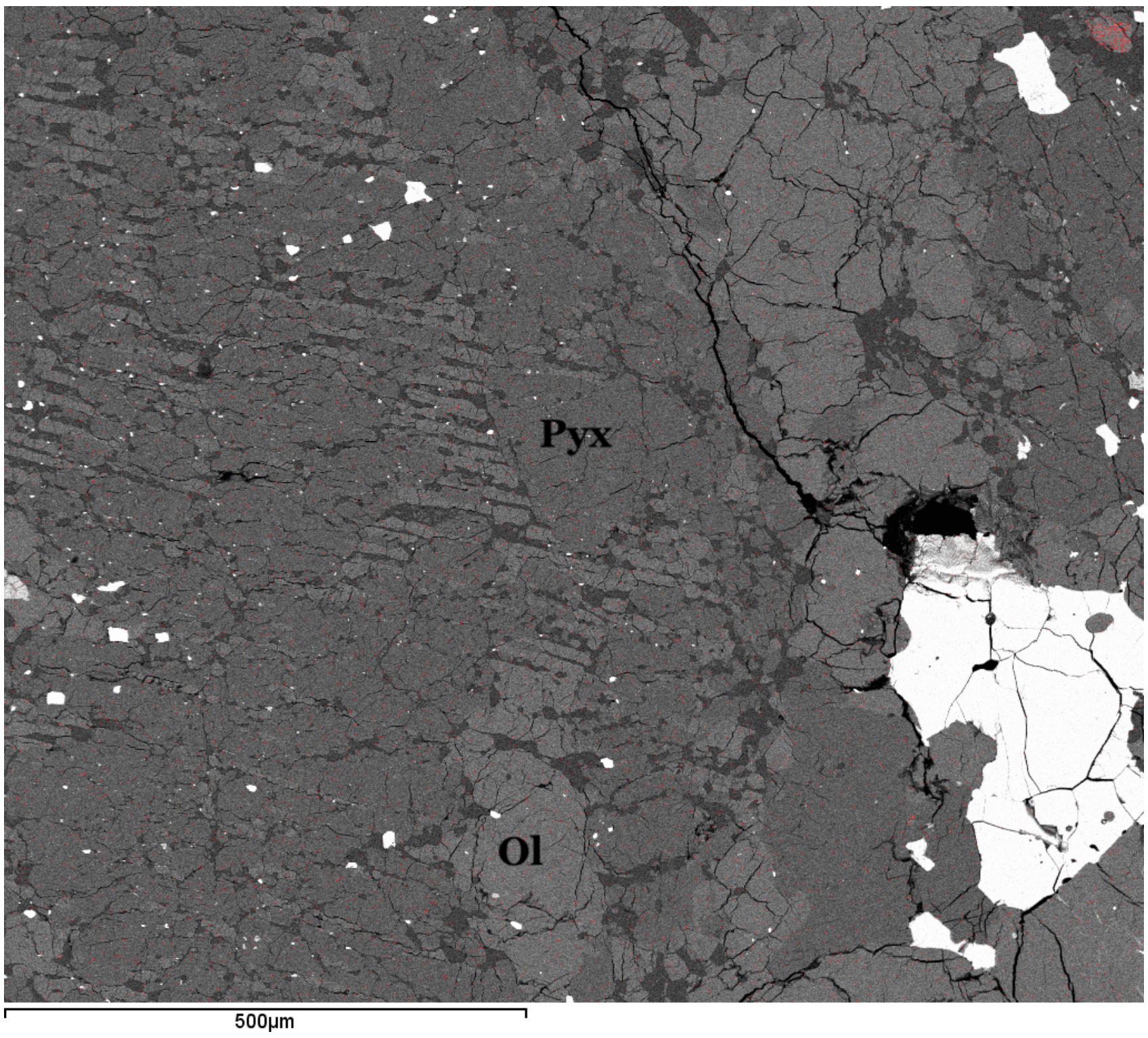

Figure S4: Backscattered electron image 157 of Lithology A – textural evidence for pyroxene replacement of olivine. Mineral abbreviations: Olivine = Ol; Pyroxene = Pyx.

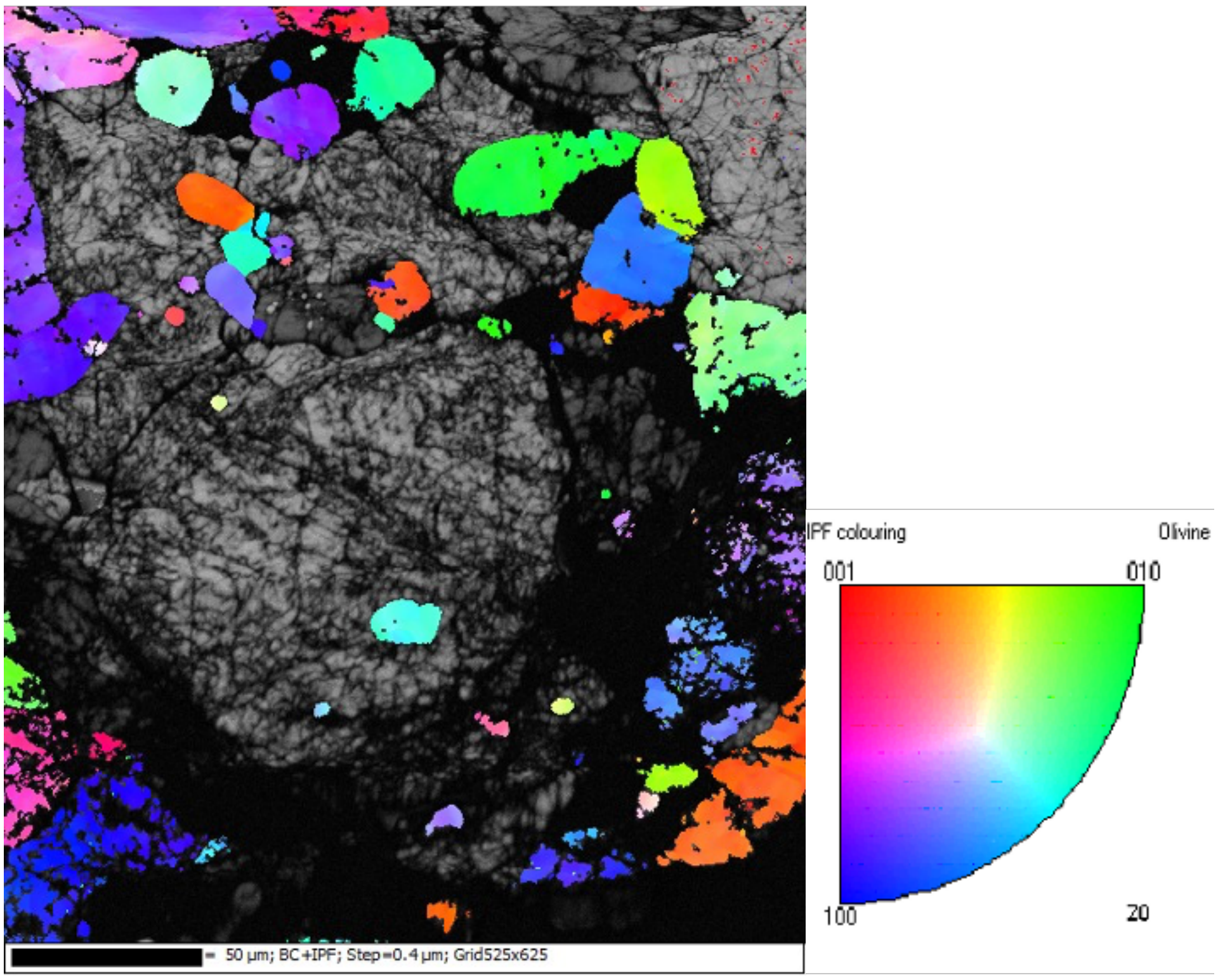

Figure S5A: Inverse Pole Figure (IPF) map of olivine surrounding and included within MERR-280-A. Individual olivine grains show random orientations, according to IPF maps. Several of these rounded olivine grains show identical orientations yet are now separated from one another via (secondary – see main text) phosphate growth. This textural relationship strongly suggests that a replacement reaction of olivine to form phosphate has taken place. Data are for the orthorhombic structure of olivine of mmm crystal class, in an Equal Area lower hemispheres projection.

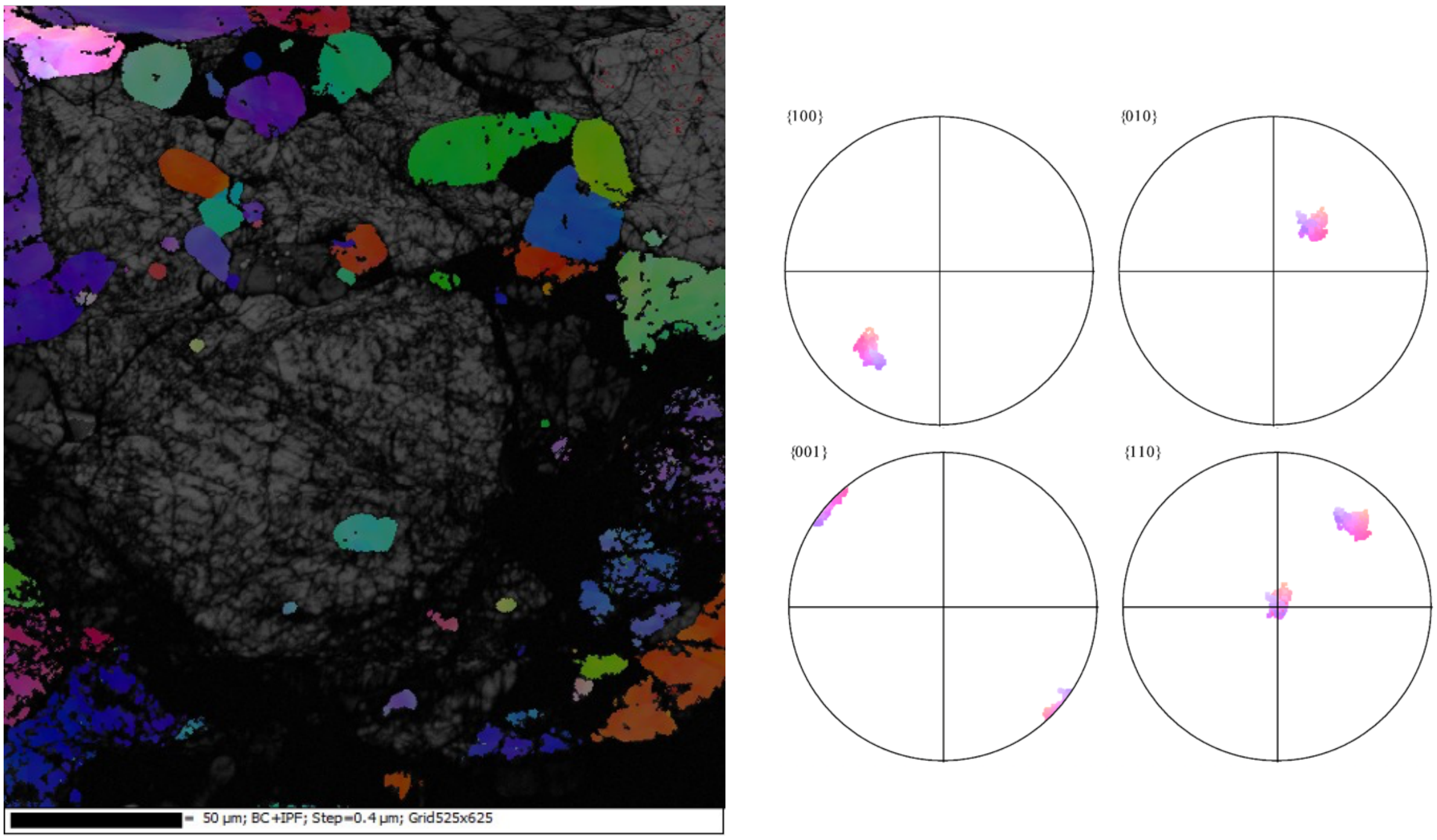

Figure S5-B(i): Microtexture and internal deformation of selected (highlighted) olivine grain revealed by IPF map and related pole figure show compact internal structure of olivine in contact with merrillite, and minimal crystal-plastic deformation. Data are for orthorhombic structure of olivine of mmm crystal class (Hermann-Mauguin notation), in an Equal Area lower hemispheres projection. There are 2745 data points.

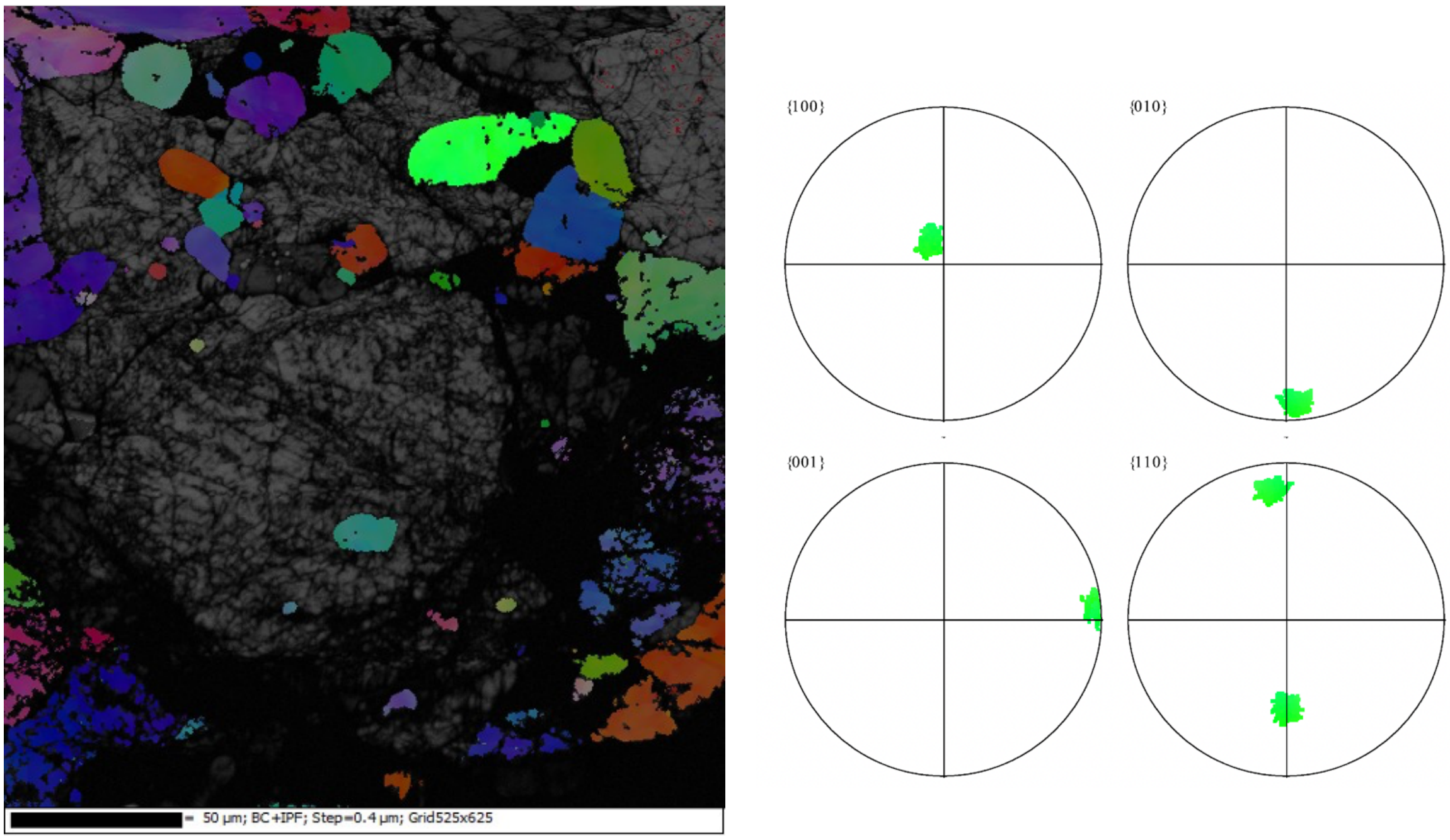

Figure S5-B(ii): Microtexture and internal deformation of selected (highlighted) olivine grain revealed by IPF map and related pole figure show compact internal structure of olivine in contact with merrillite, and minimal crystal-plastic deformation. Data are for orthorhombic structure of olivine of mmm crystal class (Hermann-Mauguin notation), in an Equal Area lower hemispheres projection. There are 4078 data points.

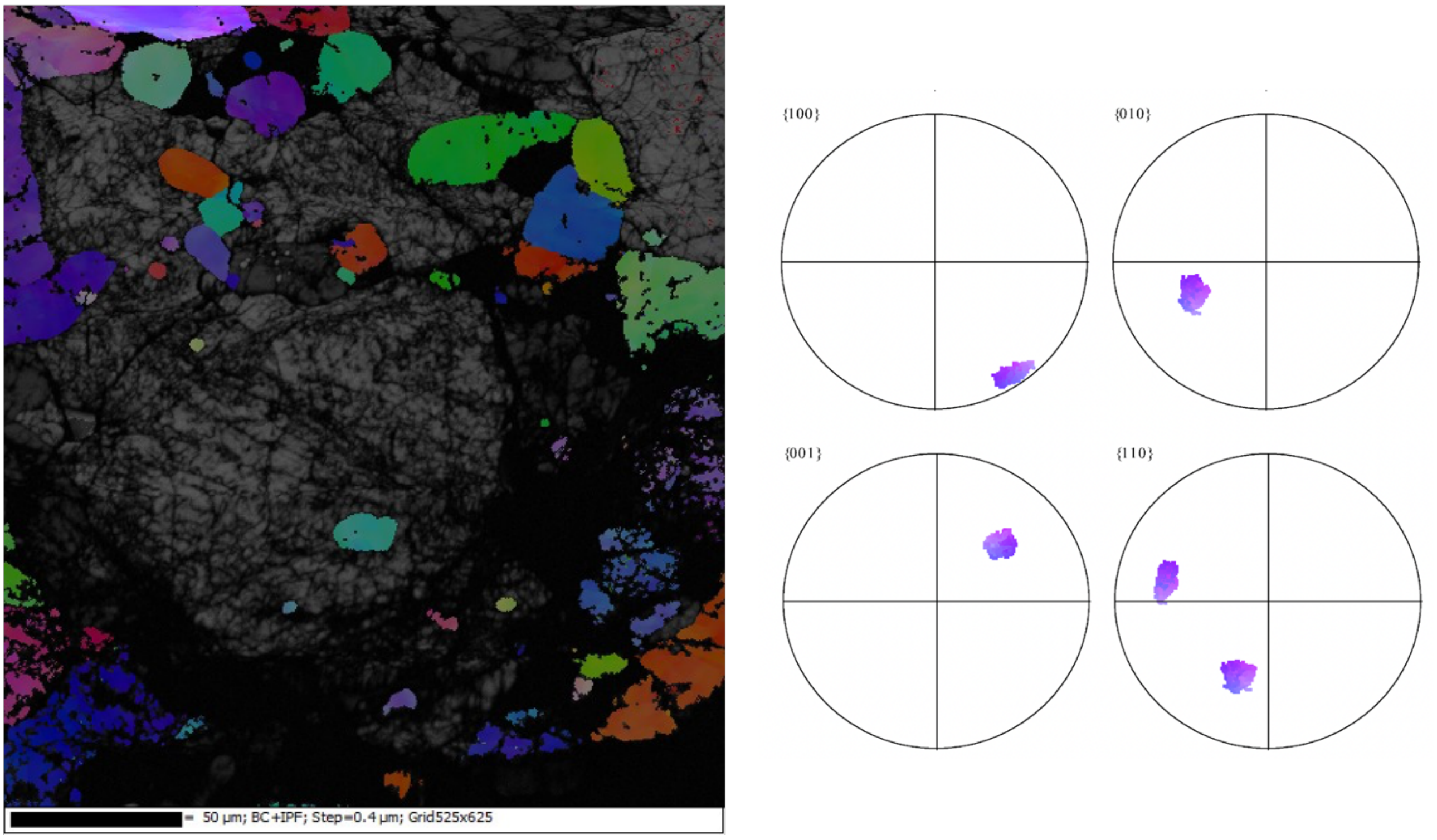

Figure S5-B(iii): Microtexture and internal deformation of selected (highlighted) olivine grain revealed by IPF map and related pole figure show compact internal structure of olivine in contact with merrillite, and minimal crystal-plastic deformation. Data are for orthorhombic structure of olivine of mmm crystal class (Hermann-Mauguin notation), in an Equal Area lower hemispheres projection. There are 2775 data points.

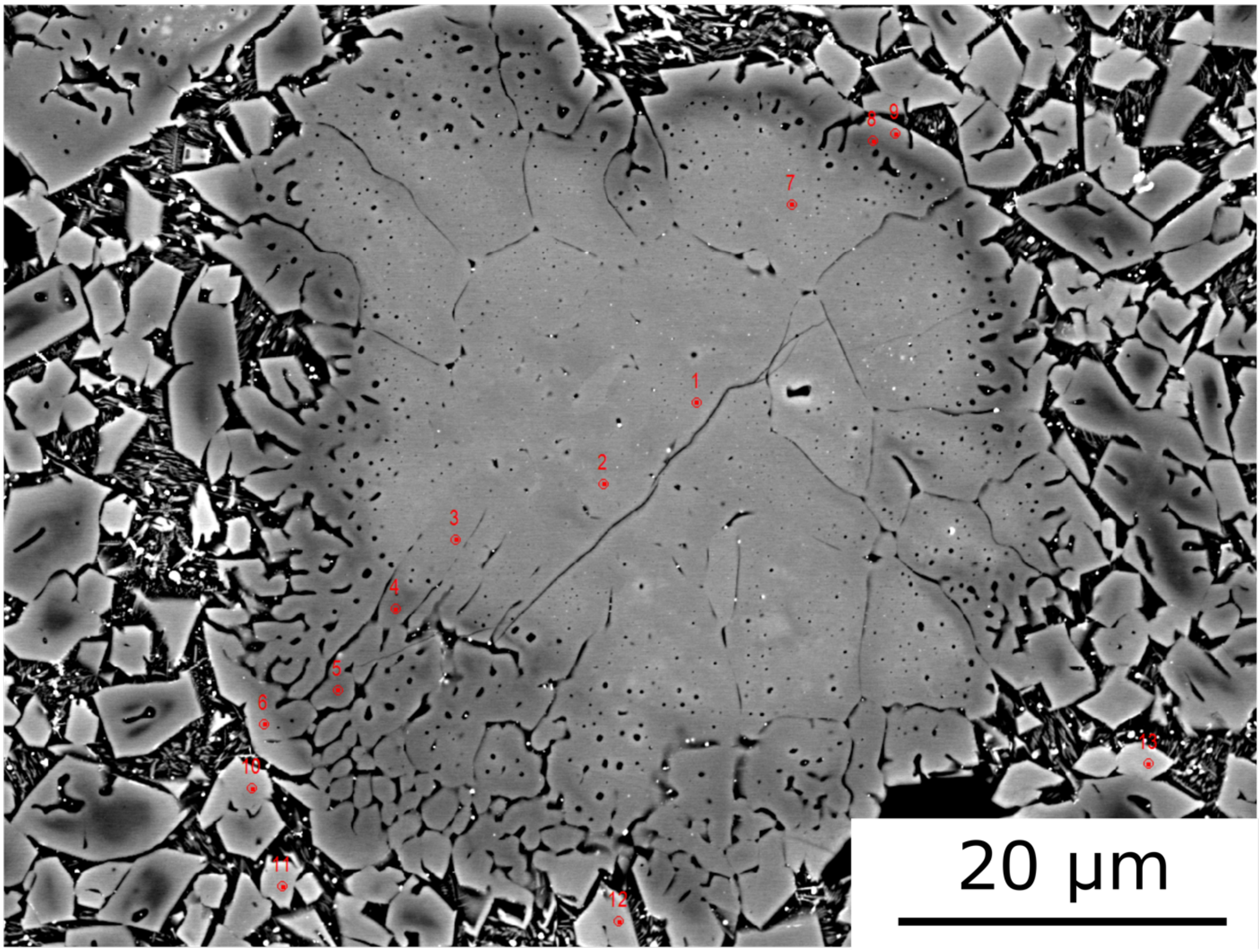

Figure S6: Correlation of EPMA spot analyses with BSE image of olivine grain B12 in the melt lithology. Data for each point can be found in File SI-2.

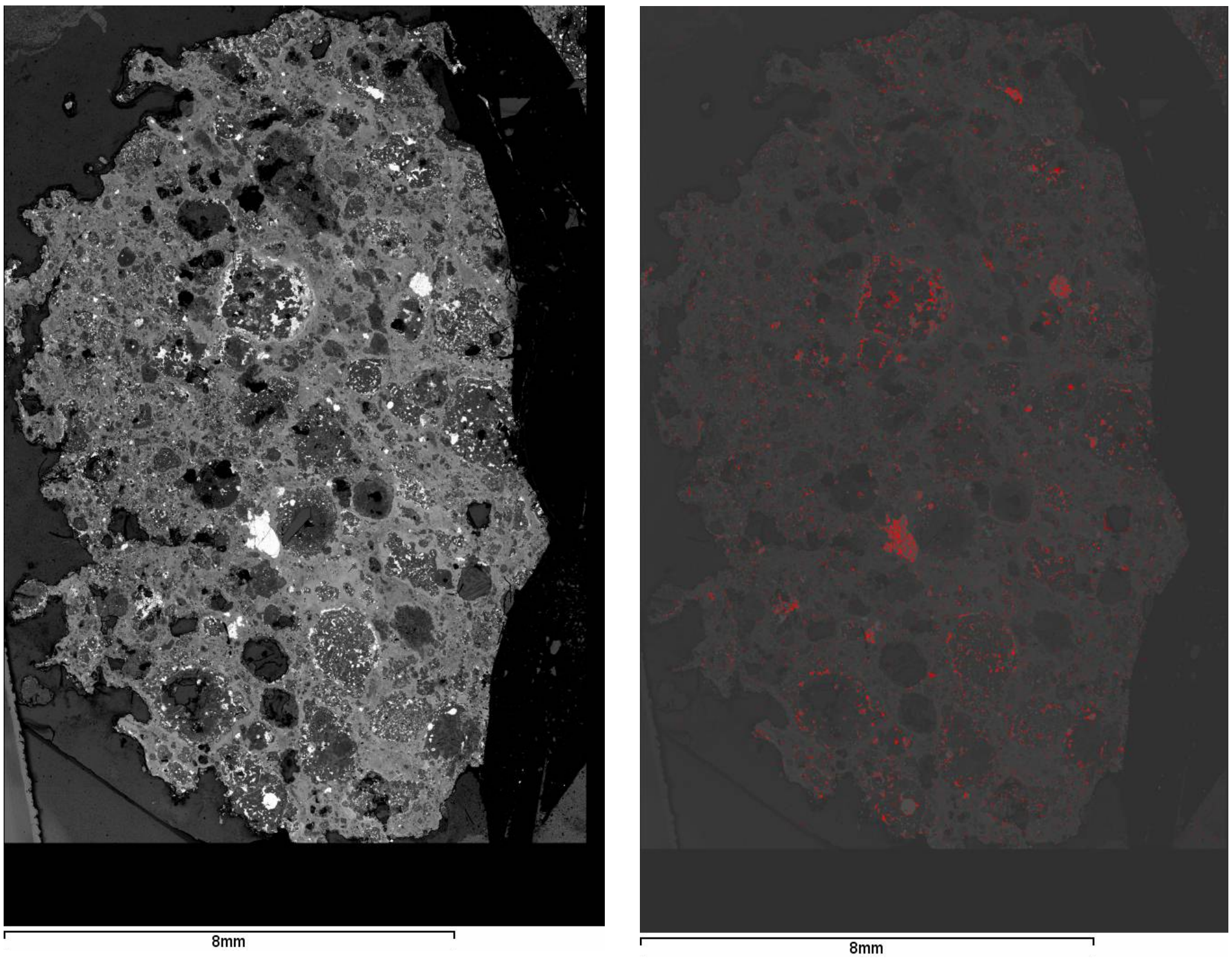

Figure S7A: Backscattered electron image of a section of Vigarano (NHM London loan).
Figure S7B: SEM + P EDS hotspot map of Vigarano, revealing P association dominantly with metal and silicate grains.

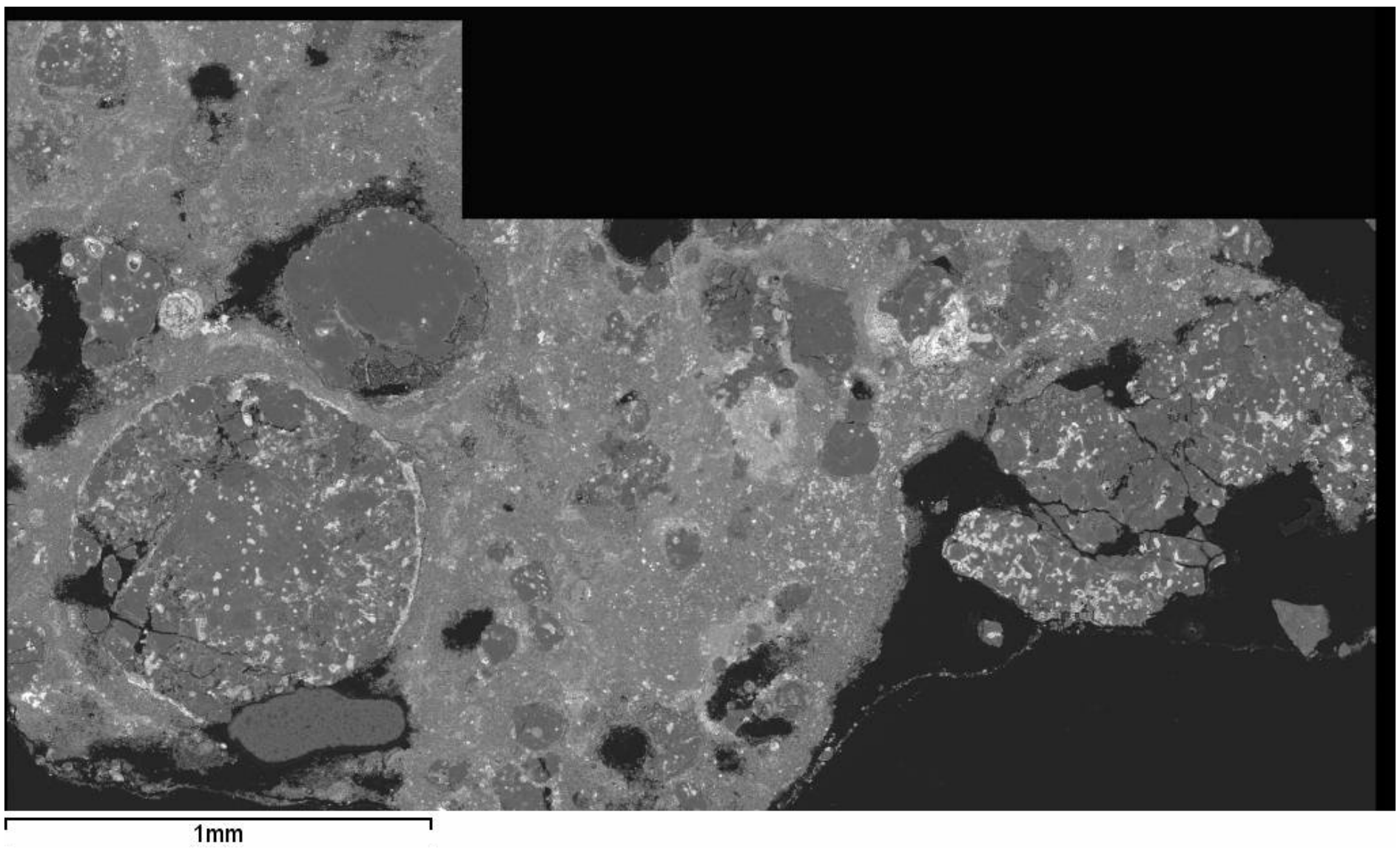

Figure S8A: Backscattered electron image of a section of Ningqiang (NHM London loan)

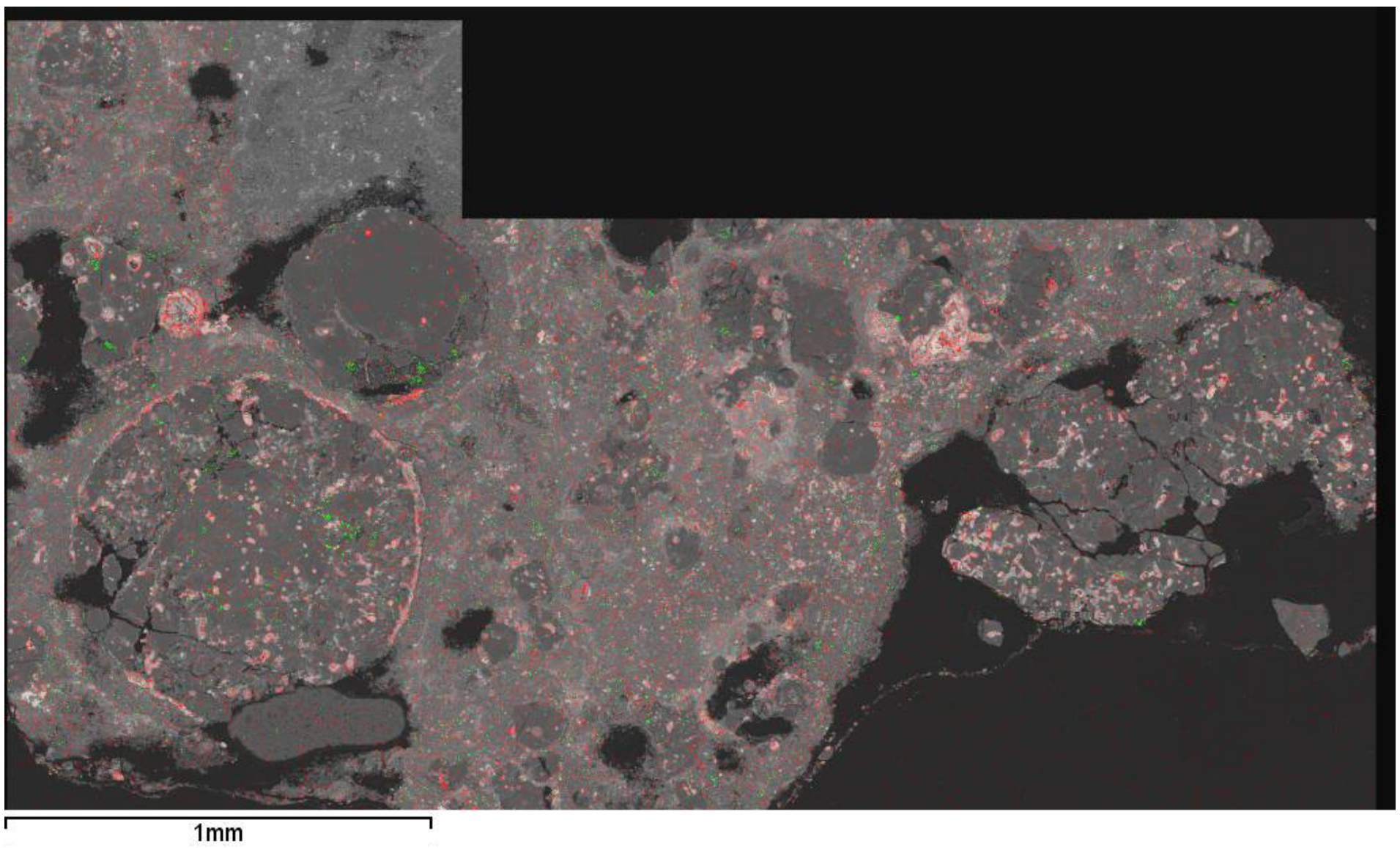

Figure S8B: SEM + P (red) and Cl (green) EDS hotspot map of Ningqiang, revealing P association dominantly with metal and silicate grains.

**Table S1:** Phase associations of discrete phosphate grains in Chelyabinsk. Individual grains of either merrillite or apatite are listed in the 'grain' column. The use of '1' indicates that a phosphate grain is observed to share a grain boundary with a particular phase. '0' indicates no observed association.

| SECTION A<br>LIGHT LITHOLOGY | | | | | | | | | | |
|---|---|---|---|---|---|---|---|---|---|---|
| **Merrillite** | **Textures/Associations** | | | | | | **General Setting** | | | |
| **Grain** | Olivine | Pyroxene | Sulphides | Metal | Void(s) | Plagioclase | SMV-proximal | Matrix | Chondrule | |
| | | | | | | | | | Rim | Interior |
| **26(1)** | 1 | 1 | 0 | 0 | 0 | 1 | 0 | 1 | 0 | 0 |
| **28(1)** | 1 | 0 | 0 | 0 | 0 | 0 | 0 | 1 | 0 | 0 |
| **40(1)** | 1 | 1 | 1 | 0 | 0 | 1 | 0 | 1 | 0 | 0 |
| **42(1)** | 1 | 0 | 0 | 0 | 0 | 0 | 0 | 1 | 0 | 0 |
| **45(1)** | 1 | 1 | 1 | 0 | 0 | 1 | 0 | 1 | 0 | 0 |
| **45(2)** | 1 | 0 | 0 | 0 | 0 | 1 | 0 | 1 | 0 | 0 |
| **45(3)** | 0 | 1 | 0 | 0 | 0 | 0 | 0 | 1 | 1 | 0 |
| **53(1)** | 1 | 0 | 0 | 0 | 0 | 1 | 0 | 1 | 0 | 0 |
| **53(2)** | 1 | 1 | 0 | 0 | 0 | 0 | 0 | 1 | 0 | 0 |
| **53(3)** | 1 | 0 | 0 | 0 | 0 | 1 | 0 | 1 | 0 | 0 |
| **54(1)** | 0 | 1 | 0 | 0 | 0 | 0 | 0 | 0 | 0 | 1 |
| **70(1)** | 1 | 1 | 0 | 0 | 0 | 1 | 0 | 1 | 0 | 0 |
| **71(1)** | 1 | 0 | 0 | 0 | 0 | 1 | 0 | 1 | 0 | 0 |
| **71(2)** | 1 | 1 | 0 | 0 | 0 | 1 | 0 | 1 | 0 | 0 |
| **71(3)** | 1 | 1 | 0 | 0 | 0 | 1 | 0 | 1 | 0 | 0 |
| **71(4)** | 1 | 1 | 0 | 0 | 0 | 1 | 0 | 1 | 0 | 0 |
| **72(1)** | 1 | 0 | 0 | 0 | 0 | 0 | 0 | 1 | 0 | 0 |
| **73(1)** | 1 | 1 | 0 | 0 | 0 | 1 | 0 | 1 | 1 | 0 |
| **73(2)** | 1 | 0 | 0 | 0 | 0 | 1 | 0 | 1 | 1 | 0 |
| **78(1)** | 0 | 1 | 0 | 0 | 0 | 0 | 0 | 1 | 0 | 0 |
| **78(2)** | 0 | 1 | 0 | 0 | 0 | 1 | 0 | 1 | 0 | 0 |
| **78(3)** | 1 | 1 | 0 | 0 | 0 | 1 | 0 | 1 | 0 | 0 |
| **78(4)** | 1 | 0 | 0 | 0 | 0 | 1 | 0 | 1 | 0 | 0 |
| **78(5)** | 1 | 0 | 0 | 0 | 0 | 1 | 0 | 1 | 0 | 0 |
| **80(1)** | 1 | 0 | 0 | 0 | 0 | 1 | 1 | 1 | 1 | 0 |
| **80(2)** | 1 | 0 | 0 | 0 | 0 | 1 | 1 | 0 | 1 | 0 |
| **80(3)** | 1 | 1 | 0 | 0 | 0 | 0 | 1 | 0 | 1 | 0 |
| **86(1)** | 1 | 0 | 0 | 1 | 0 | 0 | 0 | 1 | 0 | 0 |

**Table S1 (continued) :** Phase associations of discrete phosphate grains in Chelyabinsk. Individual grains of either merrillite or apatite are listed in the 'grain' column. The use of '1' indicates that a phosphate grain is observed to share a grain boundary with a particular phase. '0' indicates no observed association.

| SECTION A LIGHT LITHOLOGY | | | | | | | | | | |
|---|---|---|---|---|---|---|---|---|---|---|
| **Merrillite** | **Textures/Associations** | | | | | | **General Setting** | | | |
| **Grain** | Olivine | Pyroxene | Sulphides | Metal | Void(s) | Plagioclase | SMV-proximal | Matrix | Chondrule | |
| | | | | | | | | | Rim | Interior |
| **87(1)** | 1 | 1 | 0 | 0 | 0 | 1 | 0 | 0 | 1 | 0 |
| **165(1)** | 1 | 1 | 0 | 0 | 0 | 1 | 0 | 0 | 1 | 0 |
| **229(1)** | 1 | 0 | 0 | 1 | 0 | 1 | 0 | 1 | 0 | 0 |
| **229(2)** | 1 | 1 | 0 | 0 | 0 | 1 | 0 | 1 | 0 | 0 |
| **230(2)** | 0 | 1 | 1 | 0 | 0 | 1 | 0 | 1 | 0 | 0 |
| **231(1)** | 1 | 1 | 1 | 0 | 0 | 1 | 0 | 1 | 0 | 0 |
| **253(1)** | 1 | 0 | 0 | 0 | 0 | 1 | 0 | 0 | 0 | 1 |
| **253(2)** | 1 | 0 | 0 | 0 | 0 | 1 | 0 | 0 | 1 | 0 |
| **253(3)** | 1 | 0 | 1 | 0 | 0 | 1 | 0 | 0 | 1 | 0 |
| **253(4)** | 1 | 0 | 0 | 0 | 0 | 1 | 0 | 0 | 1 | 0 |
| **253(5)** | 1 | 0 | 0 | 0 | 0 | 1 | 0 | 0 | 1 | 0 |
| **253(6)** | 1 | 0 | 0 | 0 | 0 | 1 | 0 | 0 | 1 | 0 |
| **262(2)** | 1 | 1 | 0 | 0 | 0 | 1 | 0 | 1 | 0 | 0 |
| **280(1)** | 1 | 1 | 0 | 0 | 0 | 1 | 0 | 1 | 0 | 0 |
| **% Of Total Merrillites** | 0.88 | 0.52 | 0.12 | 0.05 | 0 | 0.78 | 0.07 | 0.74 | 0.31 | 0.05 |

**Table S1 (continued) :** Phase associations of discrete phosphate grains in Chelyabinsk. Individual grains of either merrillite or apatite are listed in the 'grain' column. The use of '1' indicates that a phosphate grain is observed to share a grain boundary with a particular phase. '0' indicates no observed association.

| SECTION A LIGHT LITHOLOGY | | | | | | | | | | |
|---|---|---|---|---|---|---|---|---|---|---|
| **Apatite** | **Textures/Associations** | | | | | | **General Setting** | | | |
| **Grain** | Olivine | Pyroxene | Sulphides | Metal | Void(s) | Plagioclase | SMV-proximal | Matrix | Chondrule | |
| | | | | | | | | | Rim | Interior |
| **24(1)** | 0 | 0 | 0 | 0 | 1 | 1 | 0 | 1 | 0 | 0 |
| **24(2)** | 0 | 1 | 0 | 0 | 1 | 1 | 0 | 1 | 0 | 0 |
| **24(3)** | 1 | 1 | 0 | 0 | 0 | 1 | 0 | 1 | 0 | 0 |
| **24(4)** | 1 | 0 | 0 | 0 | 0 | 1 | 0 | 1 | 0 | 0 |
| **38(1)** | 1 | 0 | 1 | 0 | 0 | 0 | 0 | 1 | 0 | 0 |
| **39(1)** | 1 | 0 | 1 | 1 | 0 | 1 | 0 | 1 | 0 | 0 |
| **39(2)** | 0 | 1 | 0 | 0 | 0 | 0 | 0 | 1 | 0 | 0 |
| **39(3)** | 0 | 1 | 0 | 0 | 0 | 0 | 0 | 1 | 0 | 0 |
| **39(4)** | 0 | 1 | 0 | 0 | 0 | 0 | 0 | 1 | 0 | 0 |
| **39(5)** | 0 | 1 | 0 | 0 | 0 | 0 | 0 | 1 | 0 | 0 |
| **41(1)** | 1 | 0 | 0 | 0 | 1 | 1 | 0 | 1 | 0 | 0 |
| **41(2)** | 1 | 1 | 0 | 0 | 0 | 0 | 0 | 1 | 0 | 0 |
| **44(1)** | 1 | 0 | 1 | 0 | 0 | 1 | 1 | 1 | 0 | 0 |
| **44(2)** | 1 | 0 | 1 | 0 | 0 | 1 | 1 | 1 | 0 | 0 |
| **44(3)** | 1 | 0 | 1 | 0 | 0 | 0 | 1 | 1 | 0 | 0 |
| **56(1)** | 0 | 1 | 0 | 0 | 0 | 1 | 0 | 1 | 0 | 0 |
| **61(1)** | 0 | 1 | 1 | 0 | 1 | 1 | 1 | 1 | 0 | 0 |
| **61(2)** | 0 | 1 | 0 | 0 | 0 | 1 | 1 | 1 | 0 | 0 |
| **61(3)** | 0 | 0 | 0 | 0 | 0 | 1 | 1 | 1 | 0 | 0 |
| **61(4)** | 1 | 1 | 0 | 0 | 0 | 1 | 1 | 1 | 0 | 0 |
| **71(2)** | 1 | 0 | 0 | 0 | 0 | 1 | 0 | 1 | 0 | 0 |
| **71(3)** | 1 | 0 | 0 | 0 | 0 | 1 | 0 | 1 | 0 | 0 |
| **71(4)** | 1 | 1 | 0 | 0 | 0 | 1 | 0 | 1 | 0 | 0 |
| **72(1)** | 1 | 0 | 0 | 0 | 0 | 0 | 0 | 1 | 0 | 0 |
| **75(1)** | 1 | 0 | 0 | 0 | 0 | 1 | 0 | 1 | 1 | 0 |
| **77(1)** | 1 | 0 | 0 | 0 | 0 | 1 | 0 | 1 | 0 | 0 |

**Table S1 (continued) :** Phase associations of discrete phosphate grains in Chelyabinsk. Individual grains of either merrillite or apatite are listed in the 'grain' column. The use of '1' indicates that a phosphate grain is observed to share a grain boundary with a particular phase. '0' indicates no observed association.

| **SECTION A LIGHT LITHOLOGY** | | | | | | | | | | |
|---|---|---|---|---|---|---|---|---|---|---|
| **Apatite** | **Textures/Associations** | | | | | | **General Setting** | | | |
| **Grain** | Olivine | Pyroxene | Sulphides | Metal | Void(s) | Plagioclase | SMV-proximal | Matrix | Chondrule | |
| | | | | | | | | | Rim | Interior |
| **79(1)** | 1 | 0 | 0 | 0 | 0 | 1 | 1 | 0 | 1 | 0 |
| **88(1)** | 1 | 0 | 0 | 0 | 0 | 1 | 0 | 1 | 0 | 0 |
| **91(1)** | 1 | 0 | 0 | 0 | 0 | 1 | 0 | 1 | 0 | 0 |
| **92(1)** | 1 | 1 | 1 | 1 | 0 | 1 | 0 | 1 | 0 | 0 |
| **92(2)** | 1 | 1 | 0 | 0 | 0 | 1 | 0 | 1 | 0 | 0 |
| **92(3)** | 1 | 0 | 0 | 0 | 0 | 0 | 0 | 1 | 0 | 0 |
| **95(1)** | 1 | 1 | 1 | 0 | 0 | 1 | 0 | 1 | 0 | 0 |
| **230(1)** | 1 | 0 | 0 | 0 | 0 | 1 | 0 | 1 | 0 | 0 |
| **% Of Total Apatites** | 0.71 | 0.44 | 0.23 | 0.06 | 0.12 | 0.74 | 0.24 | 0.97 | 0.06 | 0 |
| **(Combined) % of Total Phosphates** | 0.80 | 0.49 | 0.17 | 0.05 | 0.05 | 0.76 | 0.15 | 0.84 | 0.19 | 0.03 |

**Table S1 (continued) :** Phase associations of discrete phosphate grains in Chelyabinsk. Individual grains of either merrillite or apatite are listed in the 'grain' column. The use of '1' indicates that a phosphate grain is observed to share a grain boundary with a particular phase. '0' indicates no observed association.

| SECTION B - DARK LITHOLOGY | | | | | | | | | | |
|---|---|---|---|---|---|---|---|---|---|---|
| **Merrillite** | **Textures/Associations** | | | | | | **General Setting** | | | |
| **Grain** | Olivine | Pyroxene | Sulphides | Metal | Void(s) | Plagioclase | Shock Melt | Recrystallised Matrix | Chondrule | |
| | | | | | | | | | Rim | Interior |
| **_003** | 1 | 1 | 1 | 1 | 1 | 1 | 1 | 1 | 0 | 0 |
| **_006** | 1 | 1 | 1 | 1 | 1 | 1 | 0 | 1 | 0 | 0 |
| **_007** | 1 | 0 | 1 | 1 | 0 | 1 | 1 | 1 | 0 | 0 |
| **_008** | 1 | 1 | 1 | 1 | 1 | 1 | 0 | 1 | 0 | 0 |
| **_011** | 1 | 0 | 0 | 1 | 0 | 1 | 0 | 1 | 0 | 0 |
| **_021** | 1 | 0 | 1 | 1 | 1 | 1 | 1 | 1 | 0 | 0 |
| **_023** | 1 | 0 | 1 | 1 | 1 | 1 | 0 | 1 | 0 | 0 |
| **% Of Total Merrillites** | 1.00 | 0.43 | 0.86 | 1.00 | 0.71 | 1.00 | 0.43 | 1.00 | 0.00 | 0.00 |

**Table S1 (continued) :** Phase associations of discrete phosphate grains in Chelyabinsk. Individual grains of either merrillite or apatite are listed in the 'grain' column. The use of '1' indicates that a phosphate grain is observed to share a grain boundary with a particular phase. '0' indicates no observed association.

| SECTION B - DARK LITHOLOGY | | | | | | | | | | |
|---|---|---|---|---|---|---|---|---|---|---|
| **Apatite** | **Textures/Associations** | | | | | | **General Setting** | | | |
| **Grain** | Olivine | Pyroxene | Sulphides | Metal | Void(s) | Plagioclase | Shock Melt | Recrystallised Matrix | Chondrule | |
| | | | | | | | | | Rim | Interior |
| **_004** | 1 | 1 | 1 | 1 | 0 | 1 | 0 | 1 | 0 | 0 |
| **_007** | 1 | 1 | 1 | 1 | 0 | 1 | 0 | 1 | 0 | 0 |
| **_009** | 1 | 0 | 0 | 0 | 1 | 1 | 0 | 1 | 0 | 0 |
| **_010** | 1 | 0 | 0 | 1 | 0 | 1 | 0 | 1 | 0 | 0 |
| **_012** | 1 | 1 | 0 | 0 | 0 | 1 | 0 | 1 | 0 | 0 |
| **_013** | 1 | 1 | 0 | 0 | 1 | 1 | 0 | 1 | 0 | 0 |
| **_014** | 1 | 1 | 0 | 0 | 1 | 1 | 0 | 1 | 0 | 0 |
| **_015** | 1 | 1 | 1 | 1 | 1 | 1 | 0 | 1 | 0 | 0 |
| **_016** | 1 | 1 | 0 | 0 | 0 | 1 | 1 | 1 | 0 | 0 |
| **_017** | 1 | 1 | 1 | 1 | 1 | 1 | 0 | 1 | 0 | 0 |
| **_018** | 1 | 1 | 1 | 1 | 0 | 1 | 1 | 1 | 0 | 0 |
| **_20** | 1 | 0 | 1 | 1 | 0 | 1 | 1 | 1 | 0 | 0 |
| **_22** | 1 | 1 | 1 | 1 | 1 | 1 | 1 | 1 | 0 | 0 |
| **% Of Total Apatites** | 1 | 0.77 | 0.54 | 0.61 | 0.46 | 1 | 0.31 | 1 | 0 | 0 |
| **(Combined) % of Total Phosphates** | 1 | 0.65 | 0.65 | 0.75 | 0.55 | 1 | 0.35 | 1 | 0 | 0 |

**Table S2A: Shock Melt Lithology olivine microprobe analyses**

| Ol/No.Anal. Oxide in wt% | 5 | 7 | 9 | 4 | 8 | 4 | 5 | 1 | 2 | 3 | 4 | 6 | 9 | 8 | 5 |
|---|---|---|---|---|---|---|---|---|---|---|---|---|---|---|---|
| **$SiO_2$** | 38.8 | 38.2 | 37.6 | 38.6 | 40.8 | 36.7 | 37.2 | 38.6 | 37.2 | 37.3 | 37.1 | 37.7 | 40.3 | 38.1 | 37.6 |
| **$TiO_2$** | n.d. | 0.13 | 0.04 | 0.11 | 0.06 | 0.14 | 0.01 | n.d. | n.d. | n.d. | 0.02 | 0.02 | 0.08 | 0.02 | 0.10 |
| **$Al_2O_3$** | 0.50 | 0.20 | 0.10 | 0.32 | 1.19 | 0.06 | 0.02 | 0.60 | 0.05 | 0.03 | 0.03 | 0.22 | 0.77 | 0.13 | 1.71 |
| **$Cr_2O_3$** | 0.47 | 0.33 | 0.47 | 0.24 | 0.49 | 0.32 | 0.39 | 0.45 | 0.27 | 0.12 | 0.21 | 0.35 | 0.37 | 0.32 | 0.58 |
| **FeO** | 23.3 | 20.9 | 21.7 | 22.7 | 21.5 | 24.9 | 23.8 | 27.1 | 27.1 | 24.9 | 25.0 | 26.0 | 22.1 | 23.3 | 26.9 |
| **MnO** | 0.46 | 0.32 | 0.38 | 0.43 | 0.38 | 0.47 | 0.38 | 0.47 | 0.58 | 0.43 | 0.48 | 0.44 | 0.44 | 0.40 | 0.41 |
| **MgO** | 35.4 | 40.0 | 40.3 | 38.4 | 34.3 | 36.2 | 37.3 | 31.7 | 34.7 | 36.7 | 35.7 | 34.6 | 35.1 | 37.1 | 31.1 |
| **NiO** | 0.04 | 0.12 | 0.03 | 0.07 | n.d. | 0.03 | 0.02 | 0.05 | 0.02 | 0.03 | 0.02 | 0.01 | 0.03 | 0.03 | 0.08 |
| **CaO** | 1.05 | 0.25 | 0.18 | 0.30 | 1.47 | 0.14 | 0.15 | 0.90 | 0.16 | 0.11 | 0.12 | 0.40 | 0.52 | 0.17 | 0.53 |
| **$Na_2O$** | 0.05 | 0.06 | 0.04 | 0.06 | 0.09 | 0.06 | n.d. | 0.02 | 0.04 | 0.02 | 0.01 | 0.04 | 0.03 | 0.03 | 0.20 |
| **K2O** | 0.04 | 0.02 | n.d. | 0.02 | 0.09 | 0.01 | 0.02 | 0.04 | 0.01 | n.d. | 0.01 | 0.01 | 0.05 | n.d. | 0.03 |
| **$P_2O_5$** | 0.52 | 0.25 | 0.19 | 0.12 | 0.30 | 0.32 | 0.09 | 0.35 | 0.11 | 0.07 | 0.06 | 0.36 | 0.16 | 0.05 | 0.24 |
| Total | 100.7 | 100.7 | 100.9 | 101.3 | 100.7 | 99.4 | 99.4 | 100.2 | 100.2 | 99.7 | 98.8 | 100.1 | 100.0 | 99.6 | 99.5 |
| Mg# | **73.1** | **77.3** | **76.8** | **75.1** | **74.0** | **72.2** | **73.6** | **67.6** | **69.6** | **72.5** | **71.8** | **70.4** | **73.9** | **74.0** | **67.4** |
| Comment | B6-IM-ol-rim-light | B7-IM-ol3small-rim-light | B8-IM-ol4small-rim-light | B9-IM-ol-rim-light | B9-IM-ol-rim-light | B11-IM-ol-rim-light | B11-IM-ol-rim-light | B11-IM-ol-small-light | B11-IM-ol-small-light | B11-IM-ol-small-light | B11-IM-ol-small-light | B12-IM-ol-rim-light | B12-IM-ol-rim-light | B13-IM-ol-rim-light | N9834_B17-IM-ol-light |

**Table S2A: Shock Melt Lithology olivine microprobe analyses (continued)**

| Ol/No.Anal. Oxide in wt% | 2 | 4 | 6 | 8 | 3 | 7 | 2 | 3 | 4 | 5 | 8 | 7 | 3 | 4 |
|---|---|---|---|---|---|---|---|---|---|---|---|---|---|---|
| **$SiO_2$** | 38.2 | 38.4 | 39.1 | 38.7 | 38.7 | 38.5 | 39.1 | 38.9 | 39.1 | 39.1 | 38.5 | 37.9 | 37.7 | 37.6 |
| **$TiO_2$** | 0.05 | 0.06 | n.d. | n.d. | n.d. | 0.06 | 0.06 | 0.02 | 0.02 | n.d. | 0.04 | n.d. | n.d. | 0.04 |
| **$Al_2O_3$** | 0.12 | 0.12 | 0.26 | 0.15 | 0.30 | 0.17 | 0.32 | 0.09 | 0.35 | 0.43 | 0.16 | 0.42 | 2.03 | 1.96 |
| **$Cr_2O_3$** | 0.35 | 0.33 | 0.26 | 0.36 | 0.24 | 0.37 | 0.26 | 0.29 | 0.25 | 0.32 | 0.37 | 0.27 | 0.71 | 0.65 |
| **FeO** | 17.0 | 18.7 | 13.4 | 14.6 | 19.5 | 18.5 | 18.6 | 16.1 | 20.7 | 19.6 | 19.9 | 22.2 | 22.9 | 22.4 |
| **MnO** | 0.26 | 0.33 | 0.23 | 0.28 | 0.29 | 0.27 | 0.28 | 0.31 | 0.32 | 0.34 | 0.38 | 0.37 | 0.38 | 0.36 |
| **MgO** | 42.6 | 41.9 | 46.2 | 45.8 | 40.4 | 41.5 | 40.6 | 43.5 | 38.3 | 38.8 | 39.9 | 37.2 | 34.7 | 34.9 |
| **NiO** | 0.14 | 0.07 | 0.14 | 0.11 | 0.08 | 0.05 | n.d. | 0.07 | n.d. | 0.06 | 0.18 | 0.09 | 0.08 | 0.05 |
| **CaO** | 0.13 | 0.13 | 0.13 | 0.12 | 0.12 | 0.21 | 0.12 | 0.13 | 0.12 | 0.14 | 0.13 | 0.24 | 0.38 | 0.30 |
| **$Na_2O$** | 0.08 | 0.08 | 0.15 | 0.16 | 0.10 | 0.08 | 0.05 | 0.10 | 0.06 | 0.05 | 0.10 | 0.10 | 0.22 | 0.22 |
| **K2O** | 0.01 | 0.01 | n.d. | 0.01 | 0.01 | 0.02 | 0.01 | 0.01 | 0.02 | 0.03 | n.d. | 0.03 | 0.03 | 0.03 |
| **$P_2O_5$** | 0.19 | 0.13 | 0.18 | 0.15 | 0.04 | 0.20 | 0.12 | 0.09 | 0.05 | 0.11 | 0.08 | 0.08 | 0.11 | 0.16 |
| Total | 99.1 | 100.2 | 100.1 | 100.5 | 99.9 | 99.9 | 99.6 | 99.5 | 99.3 | 99.0 | 99.7 | 98.9 | 99.2 | 98.6 |
| Mg# | **81.7** | **79.9** | **86.0** | **84.9** | **78.7** | **80.0** | **79.5** | **82.8** | **76.7** | **77.9** | **78.2** | **75.0** | **73.0** | **73.5** |
| Comment | B1-IM-ol-rim-dark | B6-IM-ol-rim-dark | B7-IM-ol3small-core-dark | B8-IM-ol4small-core-dark | B9-IM-ol-rim-dark | B9-IM-ol-rim-dark | B11-IM-ol-inter-dark | B11-IM-ol-inter-dark | B12-IM-ol-inter-dark | B12-IM-ol-inter-dark | B12-IM-ol-inter-dark | B13-IM-ol-rim-dark | B17-IM-ol-dark | B17-IM-ol-dark |

**Table S2A: Shock Melt Lithology olivine microprobe analyses (continued)**

| Ol/No.Anal. Oxide in wt% | 1 | 3 | 1 | 1 | 1 | 1 | 2 | 3 | 3 | 1 | 1 | 2 | 3 |
|---|---|---|---|---|---|---|---|---|---|---|---|---|---|
| **$SiO_2$** | 36.4 | 36.7 | 37.1 | 38.0 | 36.9 | 37.0 | 36.9 | 37.3 | 37.1 | 36.4 | 37.0 | 36.4 | 37.0 |
| **$TiO_2$** | n.d. | n.d. | 0.03 | 0.06 | 0.01 | n.d. | 0.04 | n.d. | 0.01 | 0.06 | n.d. | n.d. | n.d. |
| **$Al_2O_3$** | 0.04 | 0.01 | n.d. | 0.20 | 0.04 | 0.03 | n.d. | 0.01 | n.d. | n.d. | 0.03 | n.d. | n.d. |
| **$Cr_2O_3$** | 0.04 | 0.01 | 0.06 | 0.12 | 0.03 | 0.05 | 0.07 | 0.04 | 0.14 | 0.00 | 0.03 | 0.03 | 0.04 |
| **FeO** | 26.6 | 26.6 | 26.8 | 21.2 | 27.2 | 26.1 | 27.4 | 27.1 | 26.6 | 26.8 | 26.4 | 29.2 | 27.7 |
| **MnO** | 0.38 | 0.49 | 0.45 | 0.32 | 0.47 | 0.44 | 0.45 | 0.44 | 0.44 | 0.46 | 0.43 | 0.49 | 0.43 |
| **MgO** | 36.1 | 36.3 | 35.5 | 39.0 | 34.9 | 34.6 | 34.9 | 35.1 | 34.9 | 35.1 | 35.5 | 33.3 | 35.2 |
| **NiO** | 0.01 | n.d. | 0.07 | 0.03 | n.d. | 0.01 | 0.03 | n.d. | n.d. | n.d. | n.d. | n.d. | 0.01 |
| **CaO** | 0.02 | 0.04 | 0.03 | 0.06 | 0.07 | 0.03 | 0.01 | 0.02 | 0.09 | 0.03 | 0.01 | 0.04 | 0.03 |
| **$Na_2O$** | n.d. | 0.01 | 0.00 | 0.07 | n.d. | n.d. | n.d. | n.d. | 0.02 | n.d. | 0.01 | 0.01 | n.d. |
| **K2O** | 0.01 | 0.01 | n.d. | 0.01 | n.d. | n.d. | n.d. | 0.01 | n.d. | 0.01 | n.d. | n.d. | n.d. |
| **$P_2O_5$** | 0.09 | 0.05 | n.d. | 0.04 | 0.10 | 0.01 | 0.01 | n.d. | 0.02 | n.d. | 0.02 | 0.07 | 0.05 |
| **Total** | 99.7 | 100.1 | 100.0 | 99.0 | 99.7 | 98.3 | 99.8 | 100.0 | 99.3 | 98.8 | 99.4 | 99.5 | 100.4 |
| Mg# | **70.8** | **70.9** | **70.3** | **76.7** | **69.6** | **70.2** | **69.4** | **69.7** | **70.0** | **70.0** | **70.5** | **67.1** | **69.4** |
| Comment | B1-IM-ol-core | B6-IM-ol-core | B9-IM-ol-core | B11-IM-ol-core | B12-IM-ol-core | B13-IM-ol-core | B13-IM-ol-core | B13-IM-ol-core | B16-IM-ol-core | B18-Matrix-ol-core | B20c-melt-large ol-core | B20c-melt-large ol-core | B20c-melt-large ol-core |

**Table S2B: Shock Melt Lithology pyroxene microprobe analyses**

| Opx/No.Anal. Oxide in wt% | 1 | 2 | 3 | 1 | 2 | 3 | 6 | 7 | 1 | 2 | 1 | 2 |
|---|---|---|---|---|---|---|---|---|---|---|---|---|
| **$SiO_2$** | 54.1 | 54.6 | 53.7 | 54.6 | 54.7 | 54.9 | 54.1 | 55.8 | 54.1 | 53.4 | 52.9 | 51.5 |
| **$TiO_2$** | 0.17 | 0.05 | 0.07 | 0.13 | 0.02 | 0.19 | 0.23 | 0.08 | 0.13 | 0.16 | 0.05 | n.d. |
| **$Al_2O_3$** | 0.15 | 0.26 | 0.87 | 0.06 | 0.06 | 0.20 | 0.17 | 0.44 | 0.17 | 1.49 | 2.79 | 2.87 |
| **$Cr_2O_3$** | 0.15 | 0.38 | 1.34 | 0.02 | 0.02 | 0.05 | 0.15 | 0.39 | 0.09 | 1.43 | 0.49 | 0.48 |
| **FeO** | 15.9 | 14.7 | 15.9 | 15.9 | 16.2 | 16.0 | 16.3 | 8.9 | 16.3 | 16.5 | 10.2 | 11.6 |
| **MnO** | 0.48 | 0.42 | 0.38 | 0.42 | 0.48 | 0.46 | 0.40 | 0.21 | 0.44 | 0.48 | 0.26 | 0.32 |
| **MgO** | 27.1 | 27.6 | 25.1 | 27.4 | 27.6 | 28.2 | 27.4 | 32.0 | 27.2 | 24.0 | 28.0 | 26.8 |
| **NiO** | n.d. | n.d. | 0.01 | n.d. | n.d. | n.d. | 0.07 | 0.05 | 0.02 | 0.06 | 0.05 | 0.11 |
| **CaO** | 0.73 | 0.65 | 1.20 | 0.60 | 0.80 | 0.84 | 0.71 | 0.27 | 0.58 | 2.81 | 2.04 | 2.12 |
| **$Na_2O$** | n.d. | 0.01 | 0.13 | 0.02 | 0.03 | 0.02 | 0.02 | n.d. | 0.03 | 0.13 | 0.74 | 0.91 |
| **K2O** | n.d. | n.d. | 0.02 | 0.01 | 0.01 | 0.01 | n.d. | 0.01 | n.d. | 0.03 | 0.02 | 0.04 |
| **$P_2O_5$** | 0.03 | n.d. | 0.06 | n.d. | 0.03 | n.d. | n.d. | 0.06 | n.d. | 0.05 | 0.13 | 0.19 |
| **Total** | 98.8 | 98.7 | 98.7 | 99.2 | 99.9 | 100.9 | 99.4 | 98.2 | 99.0 | 100.5 | 97.6 | 96.9 |
| Mg# | **75.3** | **77.0** | **73.8** | **75.4** | **75.3** | **75.9** | **75.0** | **86.5** | **74.8** | **72.2** | **83.1** | **80.4** |
| Comment | N9834_B4c-IM-opx-core | N9834_B4c-IM-opx-rim-dark | N9834_B4c-IM-opx-rim-light | N9834_B9-IM-opx-core | N9834_B9-IM-opx-inter | N9834_B9-IM-opx-rim | N9834_B15-IM-opx | N9834_B15-IM-opx-rim | N9834_B16-IM-opx-core | N9834_B16-IM-opx-rim | N9834_B17-IM-opx-dark | N9834_B17-IM-opx-dark |

**Table S2C: Shock Melt Lithology glass microprobe analyses**

| Gl/No.Anal. Oxide in wt% | 1 | 2 | 3 | 4 |
|---|---|---|---|---|
| **$SiO_2$** | 63.3 | 64.3 | 63.5 | 63.4 |
| **$TiO_2$** | n.d. | 0.08 | n.d. | 0.06 |
| **$Al_2O_3$** | 21.9 | 21.7 | 22.8 | 20.7 |
| **$Cr_2O_3$** | 0.02 | 0.01 | 0.02 | n.d. |
| **FeO** | 1.44 | 1.14 | 1.08 | 1.89 |
| **MnO** | 0.01 | 0.03 | 0.00 | 0.06 |
| **MgO** | 0.32 | 0.23 | 0.20 | 0.73 |
| **NiO** | 0.02 | 0.03 | 0.04 | 0.08 |
| **CaO** | 3.02 | 2.58 | 2.49 | 3.36 |
| **$Na_2O$** | 1.45 | 1.24 | 1.23 | 1.49 |
| **K2O** | 0.89 | 0.93 | 0.93 | 0.84 |
| **$P_2O_5$** | n.d. | 0.01 | 0.03 | 0.05 |
| **Total** | 92.3 | 92.3 | 92.2 | 92.7 |
| Mg# | **28.2** | **26.7** | **24.3** | **40.8** |
| Comment | N9834_B1 5-IM-pl | N9834_B1 5-IM-pl | N9834_B1 5-IM-pl | N9834_B1 5-IM-pl |

**Table S2D: Light Lithology olivine microprobe analyses**

| Ol/No.Anal. Oxide in wt% | 1 | 2 | 6 | 8 | 9 | 1 | 2 | 5 | 10 |
|---|---|---|---|---|---|---|---|---|---|
| **$SiO_2$** | 38.7 | 38.1 | 37.9 | 36.9 | 37.9 | 38.1 | 38.2 | 38.2 | 38.1 |
| **FeO** | 26.6 | 26.2 | 26.9 | 25.7 | 26.1 | 26.7 | 26.6 | 26.4 | 26.7 |
| **MnO** | 0.43 | 0.47 | 0.45 | 0.40 | 0.43 | 0.45 | 0.45 | 0.44 | 0.44 |
| **MgO** | 36.4 | 36.1 | 35.7 | 34.9 | 36.0 | 36.2 | 36.2 | 36.0 | 36.1 |
| **CaO** | 0.02 | 0.02 | 0.04 | 0.05 | 0.11 | 0.01 | 0.01 | 0.02 | 0.02 |
| **$P_2O_5$** | 0.05 | 0.06 | 0.03 | 0.03 | 0.03 | 0.01 | 0.01 | n.d. | n.d. |
| **Total** | 102.2 | 101.0 | 101.0 | 98.0 | 100.6 | 101.5 | 101.4 | 101.1 | 101.3 |
| Comment | Section A - light lithology - frame 165 | Section A - light lithology - frame 165 | Section A - light lithology - frame 165 | Section A - light lithology - frame 165 | Section A - light lithology - frame 165 | Section A - light lithology - frame 267 | Section A - light lithology - frame 267 | Section A - light lithology - frame 267 | Section A - light lithology - frame 267 |